\documentclass[aps,preprint,superscriptaddress,nofootinbib,longbibliography]{revtex4-1}

\usepackage{booktabs}
\usepackage{graphicx}
\usepackage{xcolor}
\usepackage{amsmath}
\usepackage{hyperref}
\usepackage{bm}
\usepackage{mathtools}
\usepackage{amssymb}
\usepackage{lipsum}
\usepackage{epstopdf}
\usepackage{slashed}
\usepackage{multirow}
\usepackage{appendix}
\usepackage{cleveref}
\usepackage{setspace}
\usepackage{subcaption}

\AtBeginDocument{
\heavyrulewidth=.08em
\lightrulewidth=.05em
\cmidrulewidth=.03em
\belowrulesep=.65ex
\belowbottomsep=0pt
\aboverulesep=.4ex
\abovetopsep=0pt
\cmidrulesep=\doublerulesep
\cmidrulekern=.5em
\defaultaddspace=.5em
}

\allowdisplaybreaks

\begin{document}
\def\qq{\langle \bar q q \rangle}
\def\uu{\langle \bar u u \rangle}
\def\dd{\langle \bar d d \rangle}
\def\sp{\langle \bar s s \rangle}
\def\GG{\langle g_s^2 G^2 \rangle}
\def\Tr{\mbox{Tr}}
\def\figt#1#2#3{
        \begin{figure}
        $\left. \right.$
        \vspace*{-2cm}
        \begin{center}
        \includegraphics[width=10cm]{#1}
        \end{center}
        \vspace*{-0.2cm}
        \caption{#3}
        \label{#2}
        \end{figure}
    }

\def\figb#1#2#3{
        \begin{figure}
        $\left. \right.$
        \vspace*{-1cm}
        \begin{center}
        \includegraphics[width=10cm]{#1}
        \end{center}
        \vspace*{-0.2cm}
        \caption{#3}
        \label{#2}
        \end{figure}
                }

\def\ds{\displaystyle}
\def\beq{\begin{equation}}
\def\eeq{\end{equation}}
\def\bea{\begin{eqnarray}}
\def\eea{\end{eqnarray}}
\def\beeq{\begin{eqnarray}}
\def\eeeq{\end{eqnarray}}
\def\ve{\vert}
\def\vel{\left|}
\def\ver{\right|}
\def\nnb{\nonumber}
\def\ga{\left(}
\def\dr{\right)}
\def\aga{\left\{}
\def\adr{\right\}}
\def\lla{\left<}
\def\rra{\right>}
\def\rar{\rightarrow}
\def\lrar{\leftrightarrow}
\def\nnb{\nonumber}
\def\la{\langle}
\def\ra{\rangle}
\def\ba{\begin{array}}
\def\ea{\end{array}}
\def\tr{\mbox{Tr}}
\def\ssp{{\Sigma^{*+}}}
\def\sso{{\Sigma^{*0}}}
\def\ssm{{\Sigma^{*-}}}
\def\xis0{{\Xi^{*0}}}
\def\xism{{\Xi^{*-}}}
\def\qs{\la \bar s s \ra}
\def\qu{\la \bar u u \ra}
\def\qd{\la \bar d d \ra}
\def\qq{\la \bar q q \ra}
\def\gGgG{\la g^2 G^2 \ra}
\def\q{\gamma_5 \not\!q}
\def\x{\gamma_5 \not\!x}
\def\g5{\gamma_5}
\def\sb{S_Q^{cf}}
\def\sd{S_d^{be}}
\def\su{S_u^{ad}}
\def\sbp{{S}_Q^{'cf}}
\def\sdp{{S}_d^{'be}}
\def\sup{{S}_u^{'ad}}
\def\ssp{{S}_s^{'??}}

\def\sig{\sigma_{\mu \nu} \gamma_5 p^\mu q^\nu}
\def\fo{f_0(\frac{s_0}{M^2})}
\def\ffi{f_1(\frac{s_0}{M^2})}
\def\fii{f_2(\frac{s_0}{M^2})}
\def\O{{\cal O}}
\def\sl{{\Sigma^0 \Lambda}}
\def\es{ &=&}
\def\ap{&\approx&}
\def\md{&\mid& }
\def\ar{&+& }
\def\ek{&-& }
\def\kek{&-& }
\def\cp{&\times& }
\def\se{&\simeq& }
\def\eqv{&\equiv&}
\def\kpm{&\pm& }
\def\kmp{&\mp& }
\def\mcdot{\cdot}
\def\erar{&\rightarrow&}


\def\simlt{\stackrel{<}{{}_\sim}}
\def\simgt{\stackrel{>}{{}_\sim}}


\title{Semileptonic $\Xi_c$ baryon decays in the light cone QCD sum rules}

\author{T.~M.~Aliev}
\email{taliev@metu.edu.tr}
\affiliation{Department of Physics, Middle East Technical University, Ankara, 06800, Turkey}

\author{S.~Bilmis}
\email{sbilmis@metu.edu.tr}
\affiliation{Department of Physics, Middle East Technical University, Ankara, 06800, Turkey}
\affiliation{TUBITAK ULAKBIM, Ankara, 06510, Turkey}

\author{M.~Savci}
\email{savci@metu.edu.tr}
\affiliation{Department of Physics, Middle East Technical University, Ankara, 06800, Turkey}

\date{\today}

\begin{abstract}
Form factors of the weak $\Xi_c \to \Xi(\Lambda)$ transitions are calculated
within the light cone QCD sum rules. The pollutions coming from the contribution of the negative parity $\Xi_c^*$ baryon is eliminated by considering the combinations of sum rules corresponding to the different Lorentz structures. Having obtained the form factors, the branching ratios of the $\Xi_c \to 
\Xi (\Lambda) \ell \nu$ decays are also calculated, and our predictions are compared with the results of other approaches as well as the measurements done by BELLE and ALICE Collaborations.

\end{abstract}

\maketitle

\section{Introduction}
The electroweak decays of the heavy flavored hadrons provide useful
information about the helicity structure of the effective Hamiltonian and
the matrix elements of the Cabibbo--Kobayashi--Maskawa matrix (CKM). These
decays are also very promising in looking for new physics searches. For these goals,  the semileptonic decays of the heavy mesons and baryons provide an ideal research area. Since the leptonic part of these
transitions is well known, all probable complications can be attributed
to the hadronic matrix element. The study of the
semileptonic $\Xi_c \to B(B=\Xi \, {\rm or}\,\Lambda) \ell \nu$
 decays that are induced by the $c \to s$ and
$c \to d$ transitions are helpful for precise determination of  the values of the CKM matrix
elements $V_{cs}$ and $V_{cd}$ .  Moreover, these decays can also be used
to test the predictions of the heavy quark effective theory.

The form factors also play a crucial role in the theoretical analysis of the  non-leptonic decays of the baryons and might also be useful for the study of CP violation.

Significant experimental progress on the semileptonic decays of
$\Xi_c$ baryon has been achieved recently. Belle Collaboration reported
the measurement of the branching ratios of the semileptonic decays
$\Xi_c^0 \to \Xi^- \ell^+ \nu$ \cite{Belle:2021crz},
\bea
\label{nolabel}
B(\Xi_c^0 \to \Xi^- e^+ \nu_e)   \es (1.72 \pm 0.10 \pm 0.12 \pm 0.50) \%~, \nnb \\
B(\Xi_c^0 \to \Xi^- \mu^+ \nu_\mu) \es (1.71 \pm 0.17 \pm 0.13 \pm 0.50) \%~. \nnb
\eea

Moreover, ALICE Collaboration has also announced the result 
for the branching ratio of the $\Xi_c^0 \to \Xi^- \ell \nu$ transition ($B = (1.8 \pm 0.2 )\%$) \cite{ALICE:2021bli}, which
agrees with BELLE's measurement within error ranges.

The semileptonic decays of the $\Xi_c$ baryon have been comprehensively studied in the framework of different approaches,
such as the light--front formalism \cite{Geng:2018plk,Geng:2019bfz}, relativistic
quark model \cite{Faustov:2019ddj}, lattice QCD \cite{Zhang:2021oja}, 3--point QCD sum rules
\cite{Zhao:2021sje}, and light cone QCD sum rules method \cite{Azizi:2011mw}.

In this study, we calculated the form factors and branching ratios of the semileptonic decay of $\Xi_c$ within the LCSR framework. It should be noted that the same channel was already studied in \cite{Azizi:2011mw}. However, the prediction of the branching ratio obtained in that study is considerably larger than the results of the other approaches as well as experimental measurements. This study also aims to understand the source of this discrepancy. In our opinion, the reason for the discrepancy can be attributed to the fact that the interpolating current for the given heavy baryon couples not only to the ground state baryon with positive parity $J^P = \frac{1}{2}^+$ but also to a heavier baryon with negative parity $J^P = \frac{1}{2}^-$.
Hence, the dispersion relation of the $\Xi_c$ baryon gets modified
when the contribution of the negative parity $\Xi_c$ baryon is taken into
account, which is $300~MeV$ heavier compared to the ground state $\Xi_c$
baryon. In the light of new experimental data, we reanalyze the semileptonic decays of $\Xi_b \rightarrow \Xi(\Lambda) l \nu$ within light-cone sum rules in detail by taking into account the contributions of $J^P ={1 \over 2}^-$ heavy baryon.

So far, the light cone sum rules (LCSR) have successfully been
applied to the wide range of problems of the hadronic physics, such as
nucleon electromagnetic form factor \cite{Braun:2006hz}, form factors and strong
coupling constants of the heavy baryons \cite{Khodjamirian:2011jp}, rare $\Lambda_b \to
N(N^\ast) \ell^+ \ell^-$ decays \cite{Aliev:2015qea}, etc.  

The paper is organized as follows. In Section~\ref{sec:2}, the light cone sum rules
(LCSR) for the $\Xi_c  \to B(\Xi~\mbox{or}~\Lambda)$
transition form factors are derived. In Section~\ref{sec:3}, the numerical analysis of the transition form
factors is performed. This section also contains our predictions on the decay widths of the $\Xi_c  \to
B \ell \nu$ transitions. Finally, we compare our results on
the branching ratios with those predicted by the other approaches.

\section{The LCSR for the $\Xi_c \to B$ transition form
factors}
\label{sec:2}
$\Xi_c \to B(\Xi~\mbox{or}~\Lambda)$
decay is induced by the $c \to s(d)$ transition. The matrix elements induced
by the vector and axial-vector transition currents are described with the
help of the three form factors,
\bea
\label{ejlf01}
\langle \Xi_c(p-q) |  \bar q \gamma_\mu  c | B(p)\rangle =
\bar {u}_{\Xi_c}(p-q) \Big[f_{1}(q^{2})\gamma_{\mu}+{i} \frac{f_{2}(q^{2})}
{m_{\Xi_c}}\sigma_{\mu\nu}q^{\nu} + \frac{f_{3}(q^{2})}{m_{\Xi_c}} q^{\mu}
\Big] u_B(p)~,\\
\label{ejlf02}
\langle \Xi_c(p-q) |  \bar q \gamma_\mu  \gamma_5 c | B(p)\rangle =
\bar {u}_{\Xi_c}(p-q) \Big[g_{1}(q^{2})\gamma_{\mu}+{i} \frac{g_{2}(q^{2})}
{m_{\Xi_c}}\sigma_{\mu\nu}q^{\nu} + \frac{g_{3}(q^{2})}{m_{\Xi_c}} q^{\mu}
\Big] \gamma_5 u_B(p)~.
\eea
The form factors responsible for the  $\Xi_c^\ast \to B$ transition can be
obtained from Eqs. (\ref{ejlf01}) and (\ref{ejlf02}) with the
replacements $f_i \to \widetilde{f}_i$, $g_i \to \widetilde{g}_i$,
inserting the Dirac matrix $\gamma_5$ after the $\Xi_c$ baryon bispinor, and
replacing $\Xi_c$ with $\Xi_c^\ast$.

In order to derive the LCSR for the form factors, we start by considering
the following correlation function(s),
\bea
\label{ejlf03}
\Pi^{V(A)}_{\mu}(p,q) =
i\int d^{4}xe^{iqx}\langle 0 | T\{\eta_{_{\Xi_{c}}} (0) J_\mu^{V(A)}(x)\}|
B(p)\rangle~,
\eea
where $\eta_{_{\Xi_{c}}}$ is the interpolating current of the $\Xi_c$ baryon,
and $J_\mu^{V(A)} = \bar{c} \gamma_\mu q (\bar{c} \gamma_\mu \gamma_5 q)$ are
the transition currents. In further calculations, we use the general form
of the interpolating current  $\Xi_c$, \cite{Bagan:1992tp}. 
\bea
\label{ejlf04}
\eta_{_{\Xi_{c}}}\es\frac{1}{\sqrt{6}}\epsilon_{abc}
\Big\{ \vphantom{\int_0^{x_2}}2 [ q^{aT}(x)Cs^{b}(x) ]\gamma_{5}c^{c}(x)
+[ q^{aT}(x)Cc^{b}(x) ]\gamma_{5}s^{c}(x) +
[ c^{aT}(x)Cs^{b}(x) ] \gamma_{5}q^{c}(x)  \nnb \\
\ar 2 \beta [ u^{aT}(x)C\gamma_{5}s^{b}(x) ]c^{c}(x) +
\beta [ q^{aT}(x)C\gamma_{5}c^{b}(x) ] s^{c}(x) +
\beta [ c^{aT}(x)C\gamma_{5}s^{b}(x) ]q^{c}(x) \Big\}~.
\eea
Here $q$ is the light quark, $C$ is the charge conjugation operator,
$a,~b$ and $c$ are the color indices,
and $\beta$ is an arbitrary parameter, and $\beta=-1$
corresponds to the Ioffe current.

To derive the LCSR for the transition form factors, we first
calculate the hadronic part of the correlation function, which is achieved
by inserting the full set of charmed--baryon states between the
interpolating current $\eta_{_{\Xi_{c}}}$ and the transition current $J_\mu$ in Eq.
(\ref{ejlf03}). Thus, the hadronic part contains the contributions  
of the lowest positive--parity $\Xi_c$, as well as its negative--parity partner
$\Xi_c^\ast$, i.e.,
\bea
\label{ejlf05}
\Pi_{\mu}^{V(A)}(p,q)=\sum_{i}\frac{\langle 0| \eta_{_{\Xi_{c}}}(0)
| \Xi_{c}^{i}(p-q,s)\rangle\langle
\Xi_{c}^{i}(p-q,s)|  \bar c \gamma_\mu q (\bar c \gamma_\mu \gamma_5q) |
B(p)\rangle}{m_{i}^{2}-(p-q)^{2}} + \cdots~,
\eea
where $\cdots$ denote the contributions of all excited and continuum
states with the quantum numbers of $\Xi_c$, and summation is performed over
the ground and first orbital excited states. The first term in the right hand side of the Eq.
(\ref{ejlf05}) describes the coupling of the $\Xi_c(\Xi_c^\ast)$ baryon with
the interpolating current $\eta_{_{\Xi_{c}}}$ which is defined as,
\bea
\label{ejlf06}
\langle 0 |  \eta_{_{\Xi_{c}}} (0) |
\Xi_c (p-q,s)\rangle \es \lambda_{_{\Xi_{c}}} u_{_{\Xi_{c}}}(p-q)~, \nnb \\
\langle 0 | \eta_{_{\Xi_{c}^\ast}} (0) |
\Xi_c^\ast(p-q,s)\rangle \es \lambda_{_{\Xi_{c}^\ast}} \gamma_5
u_{_{\Xi_{c}^\ast}}(p-q)~.
\eea
where $\lambda_{_{\Xi_{c}}} (\lambda_{_{\Xi_{c}^\ast}})$ is the residue of
the corresponding baryon.

Using the definitions of the transition form factors for the vector
transition current, and using the Dirac equation $\not\!p u_B (p) = m_B u_B(p)$
we get, 
\bea
\label{ejlf07}
\Pi_{\mu}^{V}(p,q) \es \frac{\lambda_{_{\Xi_{c}}}}{m_{_{\Xi_{c}}}^{2}-(p-q)^{2}}\Bigg\{ f_{1}(q^{2})
\Big[ 2p_{\mu}+(m_{_{\Xi_{c}}} - m_B) \gamma_{\mu}-2q_{\mu}+\gamma_{\mu}\not\!q \Big] \nnb\\
\ek \frac{f_{2}(q^{2})}{m_{_{\Xi_{c}}}} \Big[2p_{\mu}\not\!q
+(m_{_{\Xi_{c}}}^2-m_B^2) \gamma_{\mu} +
(m_{_{\Xi_{c}}}+m_B)\gamma_{\mu}\not\!q
-(m_{_{\Xi_{c}}}+m_B) q_{\mu} - q_{\mu}\not\!q\Big] \nnb\\
\ar \frac{f_{3}(q^{2})}{m_{_{\Xi_{c}}}} ( m_{_{\Xi_{c}}}+ m_B - \not\!q )
q_{\mu} \Bigg\}u_B(p) \nnb \\
\ar\frac{\lambda_{_{\Xi_{c}^\ast}}}{m_{_{\Xi_{c}^\ast}}^{2}-(p-q)^{2}}
\Bigg\{\widetilde{f}_1(q^{2})
\Big[- 2 p_{\mu} + (m_{_{\Xi_{c}^\ast}}+m_B) \gamma_{\mu} + 2q_{\mu}-\gamma_{\mu}\not\!q 
\Big] \nnb\\
\ar \frac{\widetilde{f}_2(q^{2})}{m_{_{\Xi_{c}^\ast}}} \Big[2p_{\mu}\not\!q +
(m_{_{\Xi_{c}^\ast}}^2-m_B^2) \gamma_{\mu}-(m_{_{\Xi_{c}^\ast}}-m_B)\gamma_{\mu}\not\!q +
(m_{_{\Xi_{c}^\ast}}-m_B)q_{\mu}-q_{\mu} \not\!q\Big] \nnb\\
\ar \frac{\widetilde{f}_3(q^{2})}{m_{_{\Xi_{c}^\ast}}} (
m_{\_{\Xi_{c}^\ast}}-m_B+\not\!q ) q_{\mu} \Bigg\} u_B(p)~.
\eea
$\Pi_{\mu}^{A}(p,q)$ can easily be obtained from $\Pi_{\mu}^{V}(p,q)$ by
making the replacements $\Big( f_i \to g_i;~\widetilde{f}_i \to
\widetilde{g}_i;~m_B \to -m_B\Big)$, and multiplying $\gamma_5$ matrix to the
right end, i.e.,
\bea
\Pi_{\mu}^{A}(p,q) \es\Pi_{\mu}^{V}(p,q)\Big( f_i \to g_i;~\widetilde{f}_i \to
\widetilde{g}_i;~m_B \to -m_B\Big) \gamma_5~. \nnb  
\eea

We now turn our attention to the calculation of the correlation function
(\ref{ejlf03}) for the $\Xi_c(\Xi_{c}^\ast) \to B$ transition. We take $(p-q)^2, q^2 \ll m_{_{\Xi_{c}}}^2$ to
justify the expansion of the product of the two currents in the correlation
function (\ref{ejlf03}) near the light cone $x^2\approx 0$,hence, the matrix element $\varepsilon^{abc} \lla 0 | q_\alpha^a(0)
s_\beta^b(0) s_\gamma^c(0) | B(p) \rra$ is obtained. This matrix element
is parametrized in terms of the $\Xi$ and $\Lambda$ baryon distribution
amplitudes (DAs) of a different twist. The explicit expressions of the 
$\Xi$ and $\Lambda$ baryon DAs can be found in
\cite{Liu:2009uc,Liu:2008yg,Wein:2015oqa}. The operator product expansion is obtained by
convolution of the hard--scattering amplitudes formed by the virtual
c--quark propagator and the $\Xi(\Lambda)$ baryon DAs with increasing
twists. In our calculations, we take into account all three particle
B--baryon DAs up to twist--6. However, we neglect the contributions
of the four-particle (quark and gluon) DAs.

Matching the coefficient of the relevant Lorentz structures in both
representations of the correlation function, we get the sum rules for the
transition form factors. Finally, we perform Borel transformation
over $-(p-q)^2$ in order to suppress the higher state and continuum
contributions and obtain the following 
sum rules for the form factors of the $\bar{c} \gamma_\mu s$ transition current,
\bea
\label{ejlf08}
\Pi_{1}^{B}(p,q) \es 2\lambda_{_{\Xi_{c}}}f_{1}(q^2)e^{-m_{_{\Xi_{c}}}^2/M^{2}}-2\lambda_{_{\Xi_{c}^\ast}}
\widetilde{f}_{1}(q^2)
e^{-m_{_{\Xi_{c}^\ast}}^2/M^{2}}, \nnb \\
\Pi_{2}^{B}(p,q) \es -2\lambda_{_{\Xi_{c}}}\frac{f_{2}(q^2)}{m_{_{\Xi_{c}}}}e^{-m_{_{\Xi_{c}}}^2/M^{2}}+
2\lambda_{_{\Xi_{c}^\ast}}\frac{\widetilde{f}_2(q^2)}{m_{_{\Xi_{c}^\ast}}}
e^{-m_{_{\Xi_{c}^\ast}}^2/M^{2}}, \nnb \\
\Pi_{3}^{B}(p,q) \es \lambda_{_{\Xi_{c}}}e^{-m_{_{\Xi_{c}}}^2/M^{2}}\Big(f_{1}(q^2)-
\frac{f_{2}(q^2)}{m_{_{\Xi_{c}}}}(m_{_{\Xi_{c}}}+
m_B)\Big) \nnb \\
\ar \lambda_{_{\Xi_{c}^\ast}}e^{-m_{_{\Xi_{c}^\ast}}^2/M^{2}}\Big(-\widetilde{f}_1(q^2)-
\frac{\widetilde{f}_2(q^2)}{m_{_{\Xi_{c}^\ast}}}(m_{_{\Xi_{c}^\ast}}-m_B)\Big), \nnb \\
\Pi_{4}^{B}(p,q) \es \lambda_{_{\Xi_{c}}}e^{-m_{_{\Xi_{c}}}^2/M^{2}}\Big((m_{_{\Xi_{c}}}-m_B)(f_{1}(q^2)
-\frac{f_{2}(q^2)}{m_{_{\Xi_{c}}}}
(m_{_{\Xi_{c}}}+m_B))\Big) \nnb \\
\ar  \lambda_{_{\Xi_{c}^\ast}}e^{-m_{_{\Xi_{c}^\ast}}^2/M^{2}}\Big((m_{_{\Xi_{c}^\ast}}+m_B)
(\widetilde{f}_1(q^2)+
\frac{\widetilde{f}_2(q^2)}{m_{_{\Xi_{c}^\ast}}}(m_{_{\Xi_{c}^\ast}}-m_B))\Big), \nnb \\
\Pi_{5}^{B}(p,q) \es \lambda_{_{\Xi_{c}}}e^{-m_{_{\Xi_{c}}}^2/M^{2}}\Big(-2f_{1}(q^2)+
\frac{(f_{2}(q^2)+ f_{3}(q^2))}{m_{_{\Xi_{c}}}}(m_{_{\Xi_{c}^\ast}}+m_B)\Big) \nnb \\
\ar \lambda_{_{\Xi_{c}^\ast}}e^{-m_{_{\Xi_{c}^\ast}}^2/M^{2}}\Big(2\widetilde{f}_1(q^2)+
\frac{(\widetilde{f}_2(q^2)+\widetilde{f}_3(q^2))}{m_{_{\Xi_{c}^\ast}}}
(m_{_{\Xi_{c}^\ast}}-m_B)\Big), \nnb \\
\Pi_{6}^{B}(p,q) \es \frac{\lambda_{_{\Xi_{c}}}}{m_{_{\Xi_{c}}}}e^{-m_{_{\Xi_{c}}}^2/M^{2}}
\Big(f_{2}(q^2)-f_{3}(q^2)\Big)-
\frac{\lambda_{_{\Xi_{c}^\ast}}}{m_{_{\Xi_{c}^\ast}}}e^{-m_{_{\Xi_{c}^\ast}}^2/M^{2}}
\Big(\widetilde{f}_2(q^2)-
\widetilde{f}_3(q^2)\Big).
\eea
Here,
$\Pi_{1}^{B}(p,q)$, $\Pi_{2}^{B}(p,q)$, $\Pi_{3}^{B}(p,q)$, 
$\Pi_{4}^{B}(p,q)$, $\Pi_{5}^{B}(p,q)$, and $\Pi_{6}^{B}(p,q)$
are the invariant functions for the Lorentz structures, $p_{\mu}$, 
$p_{\mu}\!\!\not\!q$,$\gamma_{\mu}\!\!\not\!q$, $\gamma_{\mu}$, $q_\mu$, and 
$q_{\mu}\!\!\not\!q$ structures, respectively.

Note that, the equations for the axial vector current, $\gamma_\mu\gamma_5$, can be obtained from Eq.
(\ref{ejlf08}) by making the following replacements
$f_i\to - g_i$, $\widetilde{f}_i\to - \widetilde{g}_i$,
$m_B\to -m_B$, and $\Pi_i^{(V)B}\to \Pi_i^{(A)B}$.

Solving the six equations given in (\ref{ejlf08}) we obtain the LCSR for the
transition form factors $f_i$, $\widetilde{f}_i$ for vector current and $g_i$, and $\widetilde{g}_i$ for axial vector
which read as:
\bea
\label{ejlf09}
f_1 \es {e^{m_{\Xi_c}^2/M^2} \over 2\lambda_{\Xi_c} (m_{_{\Xi_{c}^\ast}} + m_{_{\Xi_{c}}}) } \Big\{
(m_{_{\Xi_{c}}}+m_B) \Big[\Pi_1^{(V)B} - (m_{_{\Xi_{c}^\ast}}- m_B) \Pi_2^{(V)B} \Big] +
2 ( m_{_{\Xi_{c}^\ast}}-m_{_{\Xi_{c}}} ) \Pi_3^{(V)B} +
2 \Pi_4^{(V)B} \Big\}~, \nnb \\
f_2 \es {m_{_{\Xi_{c}}} e^{m_{\Xi_c}^2/M^2}\over 2\lambda_{\Xi_c} (m_{_{\Xi_{c}^\ast}} + m_{_{\Xi_{c}}}) }
\Big[\Pi_1^{(V)B} - (m_{_{\Xi_{c}^\ast}}- m_B) \Pi_2^{(V)B} -
2 \Pi_3^{(V)B} \Big]~, \nnb \\
f_3 \es {m_{_{\Xi_{c}}} e^{m_{\Xi_c}^2/M^2} \over 2\lambda_{\Xi_c} (m_{_{\Xi_{c}^\ast}} + m_{_{\Xi_{c}}}) }
\Big[ \Pi_1^{(V)B} + 2 (\Pi_3^{(V)B} + \Pi_5^{(V)B}) - (m_{_{\Xi_{c}^\ast}}- m_B) (\Pi_2^{(V)B}
+ 2 \Pi_6^{(V)B}) \Big]~, \nnb \\
\widetilde{f}_1 \es {- e^{m_{\Xi_c^\ast}^2/M^2} \over 2 \lambda_{\Xi_c^\ast} (m_{_{\Xi_{c}^\ast}} + m_{_{\Xi_{c}}}) } \Big\{
(m_{_{\Xi_{c}^\ast}}-m_B) \Big[\Pi_1^{(V)B} + (m_{_{\Xi_{c}}}+ m_B) \Pi_2^{(V)B} \Big] -
2 ( m_{_{\Xi_{c}^\ast}}-m_{_{\Xi_{c}}} ) \Pi_3^{(V)B} -
2 \Pi_4^{(V)B} \Big\}~, \nnb \\
\widetilde{f}_2 \es {m_{_{\Xi_{c}^\ast}} e^{m_{\Xi_c^\ast}^2/M^2} \over 2 \lambda_{\Xi_c^\ast} (m_{_{\Xi_{c}^\ast}} + m_{_{\Xi_{c}}}) }
\Big[\Pi_1^{(V)B} + (m_{_{\Xi_{c}}} + m_B) \Pi_2^{(V)B} -
2 \Pi_3^{(V)B} \Big]~, \nnb \\
\widetilde{f}_3 \es {m_{_{\Xi_{c}^\ast}} e^{m_{\Xi_c^\ast}^2/M^2} \over 2 \lambda_{\Xi_c^\ast} (m_{_{\Xi_{c}^\ast}} + m_{_{\Xi_{c}}}) }
\Big[ \Pi_1^{(V)B} + 2 (\Pi_3^{(V)B} + \Pi_5^{(V)B}) + (m_{_{\Xi_{c}}} + m_B) (\Pi_2^{(V)B}
+ 2 \Pi_6^{(V)B})  \Big]~, \nnb \\
g_1 \es {- e^{m_{\Xi_c}^2/M^2} \over 2\lambda_{\Xi_c} (m_{_{\Xi_{c}^\ast}} + m_{_{\Xi_{c}}}) } \Big\{
(m_{_{\Xi_{c}}}-m_B) \Big[\Pi_1^{(A)B} - (m_{_{\Xi_{c}^\ast}} + m_B) \Pi_2^{(A)B} \Big] +
2 ( m_{_{\Xi_{c}^\ast}}-m_{_{\Xi_{c}}} ) \Pi_3^{(A)B} + 2 \Pi_4^{(A)B}
\Big\}~, \nnb \\
g_2 \es {- m_{_{\Xi_{c}}} e^{m_{\Xi_c}^2/M^2} \over 2\lambda_{\Xi_c} (m_{_{\Xi_{c}^\ast}} + m_{_{\Xi_{c}}}) }
\Big[\Pi_1^{(A)B} - (m_{_{\Xi_{c}^\ast}} + m_B) \Pi_2^{(A)B} -
2 \Pi_3^{(A)B} \Big]~, \nnb \\
g_3 \es {- m_{_{\Xi_{c}}}e^{m_{\Xi_c}^2/M^2} \over 2\lambda_{\Xi_c} (m_{_{\Xi_{c}^\ast}} + m_{_{\Xi_{c}}}) }
\Big[ \Pi_1^{(A)B} + 2 (\Pi_3^{(A)B} + \Pi_5^{(A)B}) - (m_{_{\Xi_{c}^\ast}} + m_B) (\Pi_2^{(A)B}
+ 2 \Pi_6^{(A)B})  \Big]~, \nnb \\
\widetilde{g}_1 \es {e^{m_{\Xi_c^\ast}^2/M^2}\over 2 \lambda_{\Xi_c^\ast} (m_{_{\Xi_{c}^\ast}} + m_{_{\Xi_{c}}}) } \Big\{
(m_{_{\Xi_{c}^\ast}}+m_B) \Big[\Pi_1^{(A)B} + (m_{_{\Xi_{c}}} - m_B) \Pi_2^{(A)B} \Big] -
2 ( m_{_{\Xi_{c}^\ast}}-m_{_{\Xi_{c}}} ) \Pi_3^{(A)B} -
2 \Pi_4^{(A)B} \Big\}~, \nnb \\
\widetilde{g}_2 \es {- m_{_{\Xi_{c}^\ast}} e^{m_{\Xi_c^\ast}^2/M^2} \over 2 \lambda_{\Xi_c^\ast} (m_{_{\Xi_{c}^\ast}} + m_{_{\Xi_{c}}}) }
\Big[\Pi_1^{(A)B} + (m_{_{\Xi_{c}}} - m_B) \Pi_2^{(A)B} -
2 \Pi_3^{(A)B} \Big]~, \nnb \\
\widetilde{g}_3 \es {- m_{_{\Xi_{c}^\ast}}e^{m_{\Xi_c^\ast}^2/M^2} \over 2 \lambda_{\Xi_c^\ast} (m_{_{\Xi_{c}^\ast}} + m_{_{\Xi_{c}}}) }
\Big[ \Pi_1^{(A)B} + 2 (\Pi_3^{(A)B} + \Pi_5^{(A)B}) + (m_{_{\Xi_{c}}} - m_B) (\Pi_2^{(A)B}
+ 2 \Pi_6^{(A)B})  \Big]~. 
\eea

Few words about the theoretical calculations are in
order. The correlation function with the $\gamma_\mu$ and
$\gamma_\mu\gamma_5$ transition currents can be transformed to the following form,
\bea
\label{nolabel}
\Pi_{i}^{V(A)}[(p-q)^2,q^2]\es
\sum_{n=1,2,3}\int_{0}^{1}dx\frac{\rho_{in}^{V(A)}[x,(p-q)^2]}{\Delta^n}~, \nnb
\eea
where $\Delta=m_c^2 - x(p-q)^2 - \bar{x}q^2 +
x\bar{x} m_{_{\Xi_{c}}}^2$, and $\bar{x}=1-x$ and $\rho_{in}^{V(A)}$ are the spectral densities of the corresponding invariant functions $\Pi_{i}^{V(A)}[(p-q)^2,q^2]$. Their expressions are too lengthy, hence, we do not present here. To obtain the
relevant sum rules for the form factors, Borel transformation to the
dispersion integral representation and subtraction of continuum should be
performed. These operations can be implemented with the help of the following
replacements,
\bea
\label{ejlf10}
\int dx \frac{\rho_{i1}(x)}{\Delta} & \rightarrow & \int_{x_{0}}^{1}\frac{dx}{x} \rho_{i1}(x)
e^{\frac{-s(x)}{M^{2}}}\nnb \\
\int dx\frac{\rho_{i2}(x)}{\Delta^2} & \rightarrow & \frac{1}{M^2} \int_{x_{0}}^{1}
\frac{dx}{x^2}\rho_{i2}(x)e^{\frac{-s(x)}{M^{2}}}+ \frac{\rho_{i2}(x_{0})
e^{\frac{-s_{0}}{M^{2}}} }{m_{c}^2+x_{0}^{2}m_{_{\Xi_{c}}}^2-q^2}\nnb \\
\int dx\frac{\rho_{i3}(x)}{\Delta^3} & \rightarrow & \frac{1}{2M^4} \int_{x_{0}}^{1}
\frac{dx}{x^3}\rho_{i3}(x)e^{\frac{-s(x)}{M^{2}}}+\frac{1}{2M^2} \frac{\rho_{i3}(x_{0})
e^{\frac{-s_{0}}{M^{2}}} }{x_{0}(m_{c}^2+x_{0}^{2}m_{_{\Xi_{c}}}^2-q^2)}\nnb \\
\ek\frac{1}{2} \frac{x_{0}^2e^{\frac{-s_{0}}{M^{2}}}}{m_{c}^2+x_{0}^{2}
m_{_{\Xi_{c}}}^2-q^2}\frac{d}{dx}\Big( \frac{\rho_{i3}(x)}{x(m_{c}^2+x^{2}m_{_{\Xi_{c}}}^2-q^2)}
\Big)\Big|_{x=x_{0}}~,
\eea
and $x_{0}$ is the solution of the equation
\bea
\label{nolabel}
s_{0}=\frac{m_c^2-\bar{x}q^2+x\bar{x}m_{_{\Xi_{c}}}^2}{x}. \nnb
\eea

The expressions of the form factors involve residues of $\Xi_c$ and
$\Xi_c^\ast$ baryons. These residues can be calculated using the two-point
correlation function,
\bea
\label{nolabel06}
\Pi(q^2) \es i \int d^4x e^{iqx} \la 0 \ve \mbox{\rm T} \left\{ \eta_Q(x)
\bar{\eta}_Q(0) \right\} \ve 0 \ra~, \nnb \\
\es \Pi_1^B(q^2) \not\!q + \Pi_2^B(q^2) I~.\nnb
\eea
Following the standard sum rules methodology, namely, saturating the
correlation function with $\Xi_c$ and $\Xi_c^\ast$, and performing the Borel
transformation and continuum subtraction, we obtain,
\bea
\label{nolabel}
\Pi_1^B \es \lambda_{_{\Xi_{c}}} e^{-m_{_{\Xi_{c}}}^2/M^2} + \lambda_{_{\Xi_{c}^\ast}} 
e^{-m_{_{\Xi_{c}^\ast}}^2/M^2} ~, \nnb \\
\Pi_2^B \es \lambda_{_{\Xi_{c}}} m_{_{\Xi_{c}}} e^{-m_{_{\Xi_{c}}}^2/M^2} -  \lambda_{_{\Xi_{c}^\ast}}
m_{_{\Xi_{c}^\ast}} e^{-m_{_{\Xi_{c}^\ast}}^2/M^2} ~, \nnb
\eea
where $\lambda_{_{\Xi_{c}}}(\lambda_{_{\Xi_{c}^\ast}})$ and $m_{_{\Xi_{c}}}(m_{_{\Xi_{c}^\ast}})$ are
the residues and the masses of the $\Xi_c(\Xi_c^\ast)$ baryons, respectively. Solving
these equations, for the residue of the $\Xi_c$ baryon
we get,
\bea
\label{ejlf11}
 \lambda_{_{\Xi_{c}}} = {e^{m_{_{\Xi_{c}}}^2/M^2} \over 
m_{_{\Xi_{c}^\ast}} + m_{_{\Xi_{c}}}} \Big(m_{_{\Xi_{c}^\ast}} \Pi_1^B + \Pi_2^B
\Big)~.
\eea
The invariant functions $\Pi_1^B$ and $\Pi_2^B$ are calculated in
\cite{Aliev:2015qea}, which we will use in our numerical analysis.

\section{Numerical analysis}
\label{sec:3}
This section is devoted to the numerical analysis of the form factors derived in the previous section. The main input parameters of LCSR are the
DAs of the $\Xi$ baryon, which are calculated in \cite{Liu:2009uc,Liu:2008yg}. 

The normalization parameters of these DAs are obtained from the analysis of
the two-point sum rules (see for example \cite{Liu:2009uc,Liu:2008yg,Wein:2015oqa}), whose
values are,
\bea
\label{nolabel}
f_\Xi     \es  (9.9 \pm 0.4)\times10^{-3}~ \mbox{GeV}^2~, \nnb \\
\lambda_1 \es -(2.8 \pm 0.1)\times10^{-2}~ \mbox{GeV}^2~, \nnb\\
\lambda_2 \es  (5.2 \pm 0.2)\times10^{-2}~ \mbox{GeV}^2~, \nnb \\
\lambda_3 \es  (1.7 \pm 0.1)\times10^{-2}~ \mbox{GeV}^2~. \nnb
\eea

The numerical values of other input parameters used in the calculations are presented in Table~\ref{tab:values}.

\begin{table*}[hbt]
  \centering
  \renewcommand{\arraystretch}{1.4}
  \setlength{\tabcolsep}{7pt}
  \begin{tabular}{ccccc}
    \toprule
     Parameters             & Value         \\
    \midrule
    $m_{_{\Xi_{c}^+}}$    & $(2467.71 \pm 0.23) ~\rm{MeV}$~\cite{ParticleDataGroup:2020ssz}\\ 
    $m_{_{\Xi_{c}^0}}$    & $(2470.44 \pm 0.28) ~\rm{MeV}$~\cite{ParticleDataGroup:2020ssz}\\ 
    $m_{_{\Xi_{c}^{\ast +}}}$ & $(2791.9 \pm 0.5)~\rm{MeV}$ ~\cite{ParticleDataGroup:2020ssz} \\ 
    $m_{_{\Xi_{c}^{\ast 0}}}$ & $(2793.9 \pm 0.5)~\rm{MeV}$ ~\cite{ParticleDataGroup:2020ssz} \\ 
    $m_{_\Xi^0}$ & $(1314.86 \pm 0.20)~\rm{MeV}$~\cite{ParticleDataGroup:2020ssz}  \\ 
    $m_{_\Xi^-}$ & $(1321.71 \pm 0.07)~\rm{MeV}$~\cite{ParticleDataGroup:2020ssz}  \\ 
    $\qq(1~GeV)$ & $-(246_{-19}^{+28}~\rm{MeV})^3$ ~\cite{Khodjamirian:2011jp} \\
    $\overline{m}_c(\overline{m}_c) $     & $1.28\ \pm 0.03~\rm{GeV}$~\cite{Chetyrkin:2009fv}  \\ 
    \bottomrule
  \end{tabular}
  \caption{The values of the input parameters used in our calculations.}
  \label{tab:values}
\end{table*}

The sum rules contain three auxiliary parameters, the Borel mass $M^2$, the
continuum threshold $s_0$ and the parameter $\beta$ in the expression of the
interpolating current. According to the sum rules methodology, we should
find the working regions of these parameters, where the form factors are
practically insensitive to their variations.

The working interval of the Borel mass parameter is determined by demanding
that both the continuum and power corrections have to be sufficiently
suppressed. These requirements lead to the following working interval of the
Borel mass parameter, $M^2 = (8\pm 2)~GeV^2$. Moreover, the value of the
continuum threshold is determined by requiring that the mass sum rules
reproduce the measured mass of the lowest baryon  mass to within 10\% accuracy for
definite values of the parameter $\beta$. This leads to the threshold value, $s_0=(11\pm
1)~GeV^2$. Finally, to find the working region of $\beta$, where $\beta = \tan\theta$, we study the
dependency of mass on $\cos\theta$ at several fixed values of $M^2$ and $s_0$.
We observe that the mass exhibits good stability to the variation
of $\cos\theta$ in the interval $-1.0 < \cos\theta < -0.6$. 

It should be emphasized that, the LCSR predictions, unfortunately,
are not applicable to whole physical region 
$m_\ell^2 \le q^2 \le [m_{_{\Xi_{c}}}(m_{_{\Xi_{c}^\ast}}) -
m_{_{\Xi}}]^2$. The LCSR for the form factors are reliable only up to $q^2 \le 0.5~GeV^2$. To extend this restricted domain to
the full physical domain given above, we use the z--series parametrization
of the form factors \cite{Bourrely:2008za}, which is given as,
\bea
\label{nolabel}
z(q^2, t_{0}) = \frac{\sqrt{t_{+}-q^2}-\sqrt{t_{+}-t_0}}{\sqrt{t_+-q^2}+
\sqrt{t_+-t_0}}~, \nnb
\eea
where $t_0 = q^2_{\rm max} = [m_{_{\Xi_{c}}}(m_{_{\Xi_{c}^\ast}}) -
m_{_{\Xi}}]^2$, and  $t_+ = (m_{D_s} + m_K)^2$. 

The following parametrization 
\bea
\label{ejlf12}
f(q^2) = \frac{1}{1-q^2/(m_{\rm pole}^f)^2} \big\{ a_0^f + a_1^f\:z(q^2,t_0) +
a_2^f\:[z(q^2,t_0)]^2 \big\}~,
\eea
reproduces best fits for the form factors predicted by the LCSR.
The pole masses for the $\Xi_c \to \Xi (\Lambda)$ transition are,
\bea
\label{nolabel}
 m_{pole} = \left\{
  \begin{array}{rl}
   2.112 (2.010) ~GeV &  \!\mbox{for the form factors~~} f_1, f_2, \widetilde{g}_1,
\!\mbox{~and~} \widetilde{g}_2, \\
     2.535 (2.423)~GeV & \!\mbox{for the form factors~~} g_1, g_2,
\widetilde{f}_1, \!\mbox{~and~} \widetilde{f}_2,\\
     2.317 (2.300)~ GeV  & \!\mbox{for the form factors  } f_3~, \widetilde{g}_3~, \\
     1.969 (1.870)~ GeV  & \!\mbox{for the form factors  } g_3~.
\widetilde{f}_3 \\
  \end{array}
\right.
\eea

The values of the fit parameters
$a^f_{0}$, $a^f_1$ and $a^f_2$ for the
$\Xi_c  \to \Xi$ and $\Xi_c  \to \Lambda$
form factors are presented in Tables~\ref{tab:tab2} and ~\ref{tab:tab3},
respectively.


 \begin{table}[h]
\renewcommand{\arraystretch}{1.3}
\addtolength{\arraycolsep}{-0.5pt}
\small
$$
\begin{array}{|c|c|c|c|c|}
\hline \hline
\Xi_c \to \Xi  &     f_i(0)     &       a_0       &       a_1       &        a_2         \\  \hline
f_1                         & -0.29 \pm 0.05 & -0.70 \pm 0.08  &  12.11 \pm 1.50 &  -89.50 \pm 10.00  \\
f_2                         & -0.12 \pm 0.02 & -0.55 \pm 0.06  &  12.78 \pm 1.50 &  -95.67 \pm 11.00   \\
f_3                         & -0.49 \pm 0.10 & -1.81 \pm 0.20  &  37.68 \pm 4.00 & -263.79 \pm 15.00   \\
g_1                         & -0.22 \pm 0.04 & -0.39 \pm 0.04  &   5.52 \pm 0.60 &  -44.91 \pm 6.00   \\
g_2                         &  0.45 \pm 0.10 &  1.24 \pm 0.15  & -20.95 \pm 2.60 &  132.68 \pm 8.00   \\
g_3                         &  0.57 \pm 0.12 &  1.26 \pm 0.15  & -18.27 \pm 2.20 &  114.76  \pm 12.00 \\ \hline \hline
\end{array}
$$
\caption{Form factors of the $\Xi_c  \to \Xi$ transition}
\renewcommand{\arraystretch}{1}
\addtolength{\arraycolsep}{-1.0pt}
 \label{tab:tab2}
\end{table}

\begin{table}[h]

\renewcommand{\arraystretch}{1.3}
\addtolength{\arraycolsep}{-0.5pt}
\small
$$
\begin{array}{|c|c|c|c|c|}
\hline \hline
\Xi_c \to \Lambda  &      f_i(0)    &       a_0       &         a_1     &        a_2       \\  \hline
f_1                             & -0.36 \pm 0.06 & -0.19 \pm 0.05  &  -4.00 \pm 0.80 &  22.68 \pm 2.20  \\
f_2                             & -0.18 \pm 0.04 & -0.21 \pm 0.08  &  -0.49 \pm 0.08 &   8.77 \pm 1.00  \\
f_3                             &  0.11 \pm 0.03 &  1.17 \pm 0.12  & -19.94 \pm 2.50 &  93.66 \pm 9.40  \\
g_1                             & -0.12 \pm 0.03 &  0.26 \pm 0.04  &  -7.65 \pm 0.90 &  38.88 \pm 4.00 \\
g_2                             &  0.18 \pm 0.04 &  0.38 \pm 0.08  &  -3.10 \pm 0.50 &  10.74 \pm 1.10  \\
g_3                             & -0.05 \pm 0.01 & -0.24 \pm 0.05  &   4.28 \pm 1.00 & -22.98 \pm 2.30 \\ \hline \hline
\end{array}
$$
\caption{Form factors of the $\Xi_c \to \Lambda$ transition}
\renewcommand{\arraystretch}{1}
\addtolength{\arraycolsep}{-1.0pt}
  \label{tab:tab3}
\end{table}

\begin{figure}[!h]
\centering
\begin{subfigure}[t]{.33\textwidth}
  \centering
  \includegraphics[width=1.0\linewidth]{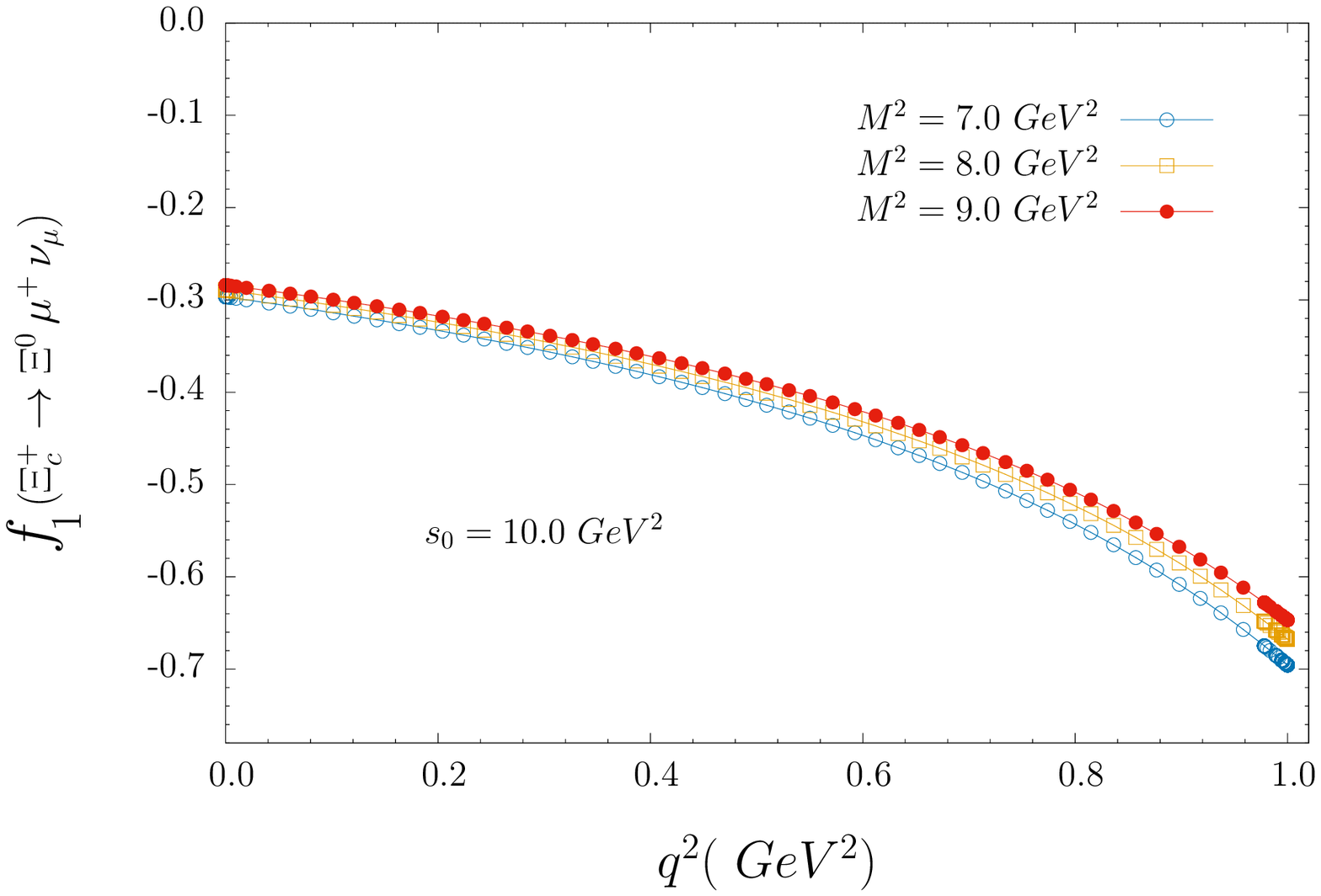}
\end{subfigure}%
\begin{subfigure}[t]{.33\textwidth}
  \centering
  \includegraphics[width=1.0\linewidth]{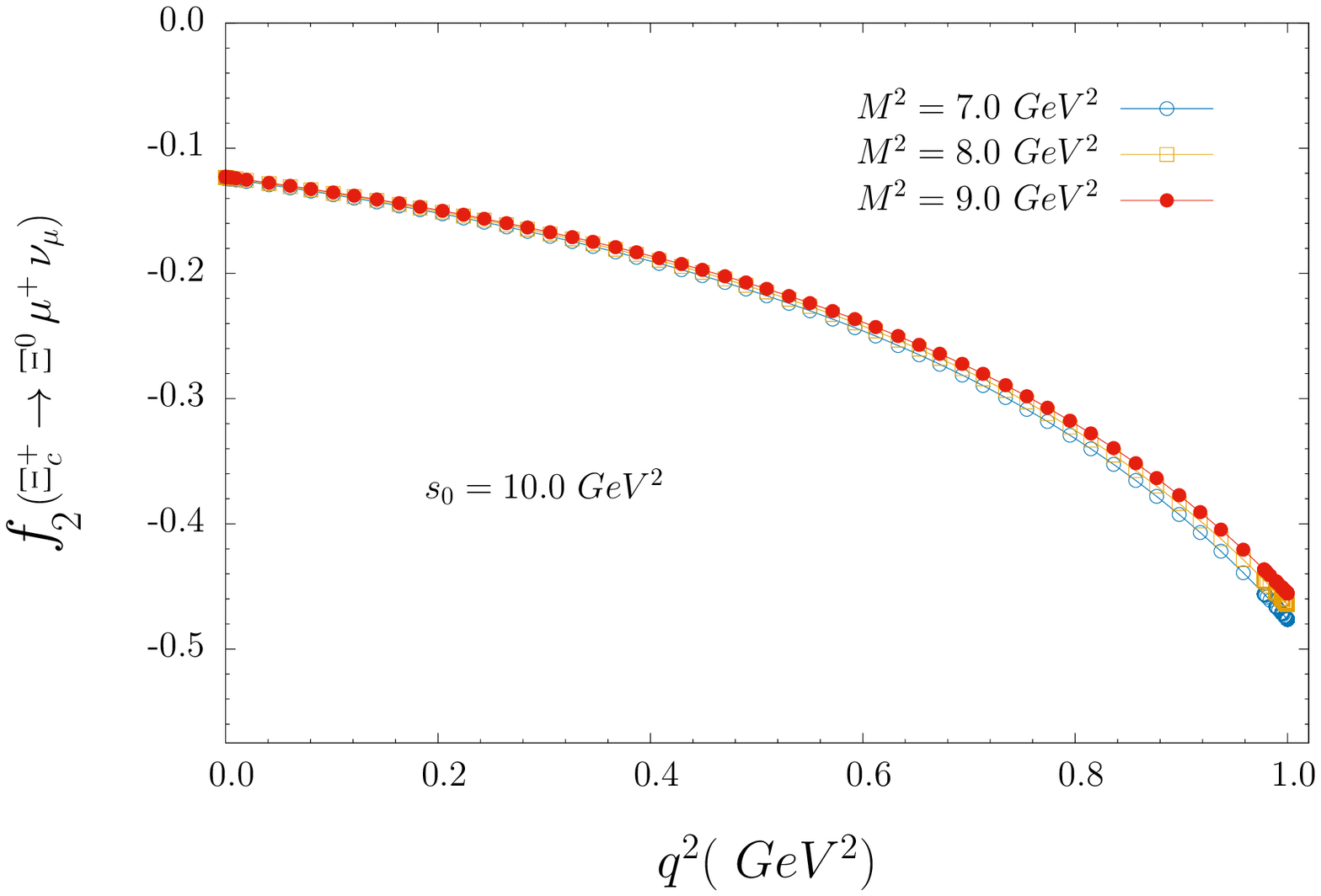}
\end{subfigure}
\begin{subfigure}[t]{.33\textwidth}
  \centering
  \includegraphics[width=1.0\linewidth]{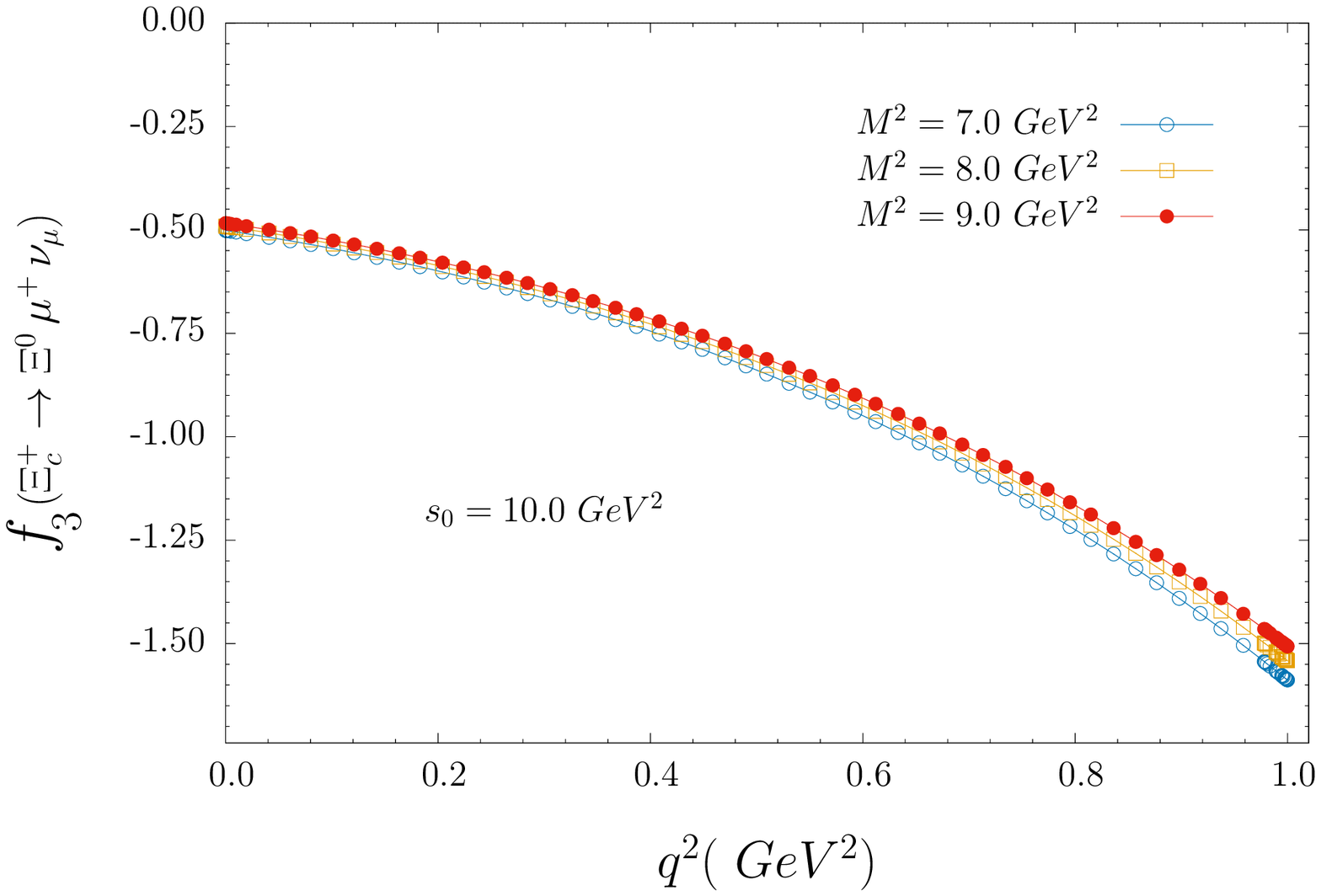}
\end{subfigure}
\caption{The dependency of the form factors $f_1$, $f_2$, and $f_3$
for the $\Xi_c^+ \rar \Xi^0 \, \mu^+ \, \nu_\mu$ transition on $q^2$,
at $s_0=10~GeV^2$, and several values of the 
Borel mass parameter $M^2$}
\label{fig:1}
\end{figure}

\begin{figure}[htb!]
\centering
\begin{subfigure}[t]{.33\textwidth}
  \centering
  \includegraphics[width=1.0\linewidth]{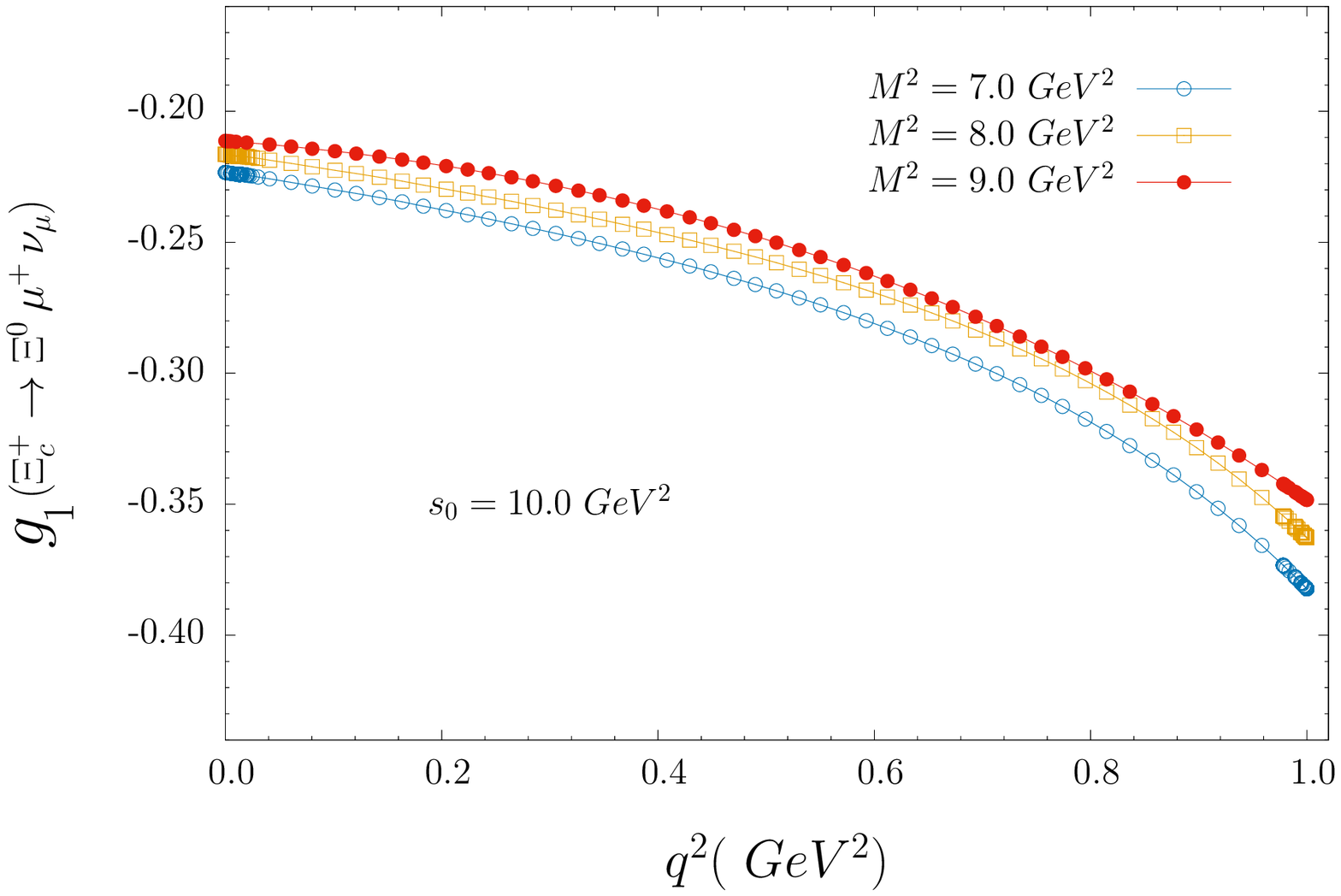}
\end{subfigure}%
\begin{subfigure}[t]{.33\textwidth}
  \centering
  \includegraphics[width=1.0\linewidth]{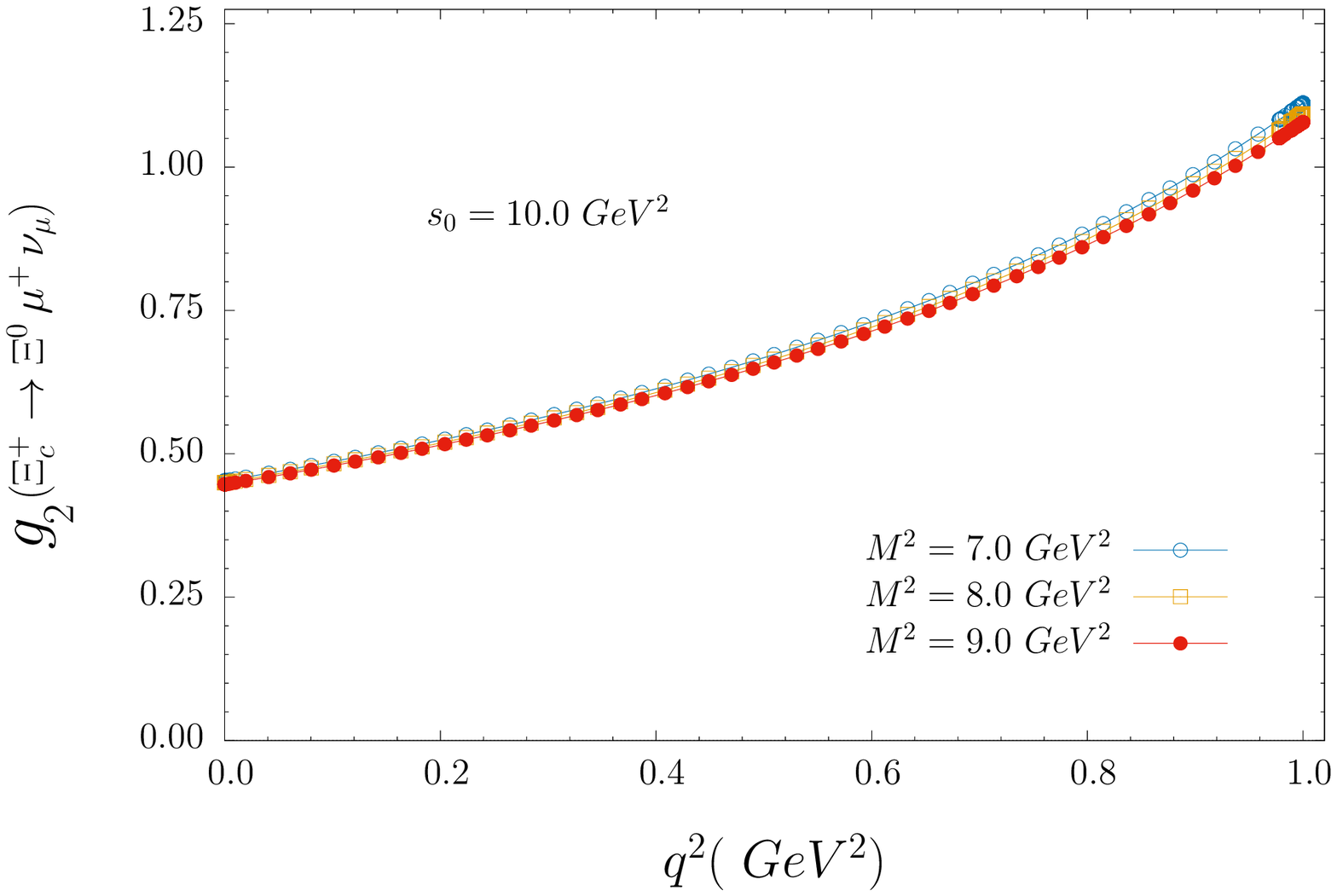}
\end{subfigure}
\begin{subfigure}[t]{.33\textwidth}
  \centering
  \includegraphics[width=1.0\linewidth]{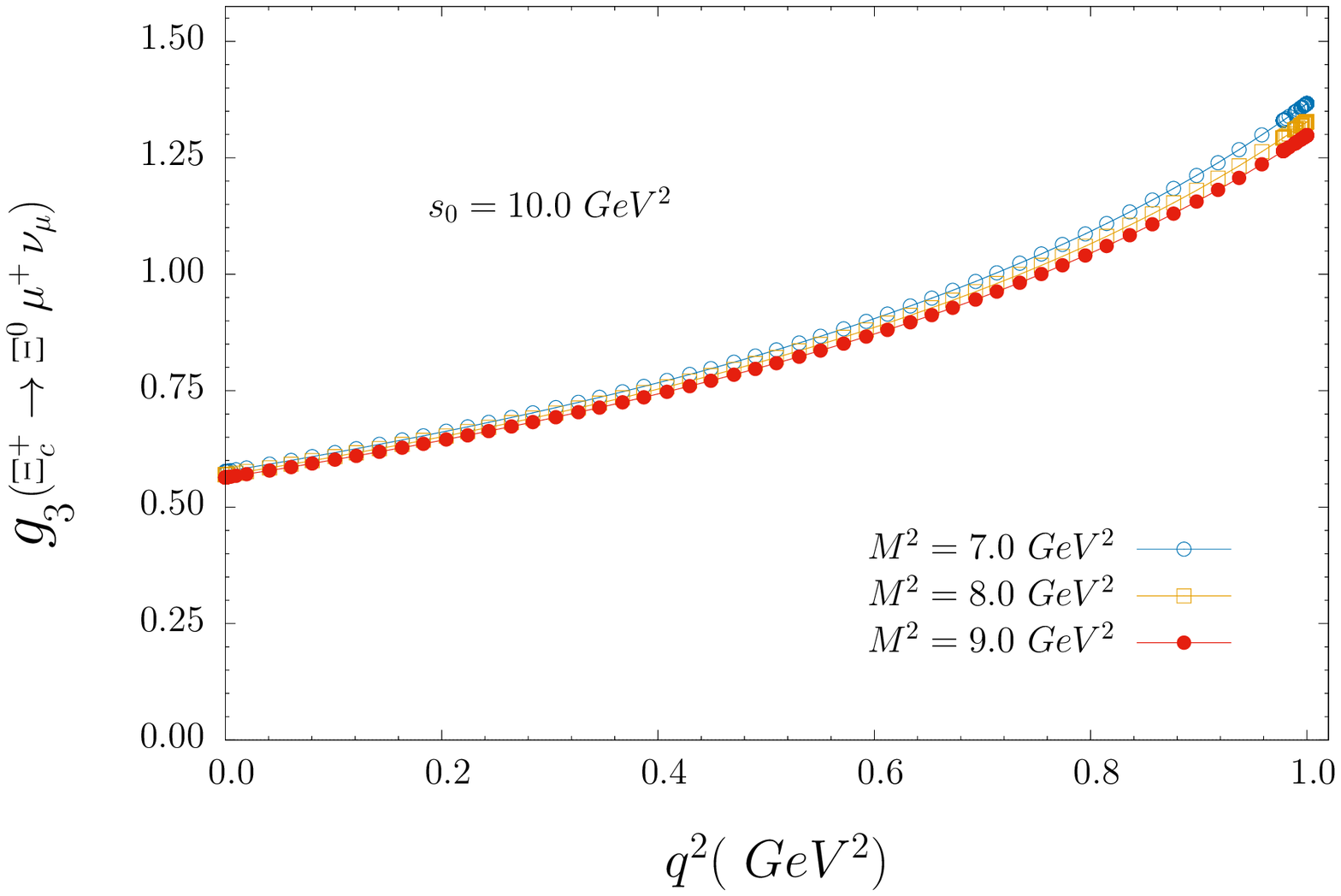}
\end{subfigure}
\caption{The dependency of the form factors $g_1$, $g_2$, and $g_3$
for the $\Xi_c^+ \rar \Xi^0 \, \mu^+ \, \nu_\mu$ transition on $q^2$,
at $s_0=10~GeV^2$, and several values of the 
Borel mass parameter $M^2$}
\label{fig:2}
\end{figure}

Dependency of the form factors $f_1$, $f_2$, $f_3$ and $g_1$, $g_2$, and $g_{3}$
on $q^2$, at the fixed value of $s_0=10.0~GeV^2$, and at several fixed
values of the Borel mass parameter $M^2$ from its working region of the
$\Xi_c \rar \Xi \, \mu \, \nu_\mu$ decay are presented in~\Cref{fig:1,fig:2},
respectively.

Having obtained the results for the form factors, we estimate the decay widths of the $\Xi_c \to B \ell \nu$ 
decays. The width of these decays can be calculated using helicity
formalism \cite{Gutsche:2015mxa}.
We choose the rest frame of $\Xi_c$ baryon, where the z-axis points along the $W_{off-shell}$ to calculate the helicity amplitudes, and we obtain
\bea
\label{ejlf??}
H_{+{1\over 2},+1}^{V(A)} \es \sqrt{2 Q_{\mp}} \left[f_1(g_1) - {m_{\pm}
\over m_{_{\Xi_{c}}}} f_2 (g_2) \right]~, \nnb\\
H_{+{1\over 2},0}^{V(A)} \es {\sqrt{Q_{\mp}}\over \sqrt{q^2}} \left[
  m_{\pm} f_1(g_1) - {q^2 \over m_{_{\Xi_{c}}}} f_2 (g_2) \right]~, \nnb \\
H_{+{1\over 2},t}^{V(A)} \es {\sqrt{Q_{\pm}}\over \sqrt{q^2}} \left[ m_{\mp} f_1 (g_1) - {q^2 \over m_{_{\Xi_{c}}}} f_3 (g_3)  \right]~, \nnb
\eea
where $m_{\pm}=  m_{_{\Xi_{c}}} \pm m_B$, and $Q_{\pm}=m_{\pm}^2-q^2$.

In these expressions, the first and second subindices  describe the helicities of the
$B$ baryon and virtual $W$, correspondingly. The amplitudes for the negative values of the
helicities can be obtained from the parity consideration, i.e.,
\bea
\label{ejlf??}
H_{-\lambda_B,-\lambda_W}^V \es   H_{\lambda_B,\lambda_W}^V~,\nnb \\
H_{-\lambda_B,-\lambda_W}^A \es - H_{\lambda_B,\lambda_W}^A~,\nnb
\eea

The total helicity amplitude is given by,
\bea
\label{ejlf??}
H_{\lambda_B,\lambda_W} = H_{\lambda_B,\lambda_W}^V -
H_{\lambda_B,\lambda_W}^A~. \nnb
\eea
Using the expressions of above-given helicity amplitudes for the differential decay widths, we obtain
\bea
\label{ejlf??}
{d \Gamma(\Xi_c \to B\ell \nu) \over dq^2} \es {G_F^2\over (2 \pi)^3} \vel
V_{cq} \ver^2 {\sqrt{\lambda(m_{_{\Xi_{c}}}^2,m_B^2,q^2)}(q^2-m_\ell^2)^2 \over
48 m_{_{\Xi_{c}}}^3 q^2} \Bigg\{
\vel H_{+{1\over 2},+1} \ver^2 +\vel H_{{-{1\over 2}},-1} \ver^2 \nnb \\
\ar \left(1+{m_\ell^2\over 2 q^2} \right) \left(\vel H_{+{1\over 2},0} \ver^2 +
\vel H_{-{1\over 2},0} \ver^2 \right)
+ {3 m_\ell^2 \over q^2} \left(  \vel H_{+{1\over 2},t} \ver^2 +\vel H_{{-{1\over 2}},t} \ver^2 \right)
\Bigg\}~,\nnb 
\eea
where $G_F$ is the Fermi constant, $V_{cq}$ is the CKM matrix element $(q=s~
{\rm or}~d)$, and
\bea
\label{ejlf??}
\lambda(m_{_{\Xi_{c}}}^2,m_B^2,q^2) = m_{_{\Xi_{c}}}^4 + m_B^4 + q^4
- 2 m^2_{{\Xi_{c}}} m_B^2 - 2 m_{_{\Xi_{c}}}^2 q^2  - 2 m_B^2 q^2~, \nnb
\eea
   
The differential decay width for the $\Xi_c^\ast \to B \ell \nu$ decay can
be obtained from $\Xi_c \to B \ell \nu$ decay by making the following
replacements, $f_1 \to -\widetilde{g}_1,~f_2 \to \widetilde{g}_2,
~g_1 \to -\widetilde{f}_1,~g_2 \to \widetilde{f}_2$, and $m_{_{\Xi_{c}}} \to 
m_{_{\Xi_{c}^\ast}}$.

Using the values of the CKM matrix elements $\vel V_{cd} \ver = 0.2211 \pm
0.0700$ and $\vel V_{cs} \ver = 0.987 \pm 0.011$ \cite{ParticleDataGroup:2020ssz} and the
$\Xi_c$ life time $\tau(\Xi_c^0) = (1.53 \pm 0.06)\times 10^{-13}~s$,
and $\tau(\Xi_c^+) = (4.56 \pm 0.05)\times 10^{-13}~s$ we can predict the
branching ratios of the corresponding semileptonic decays. Our results are
presented in Table~\ref{tab:tab4}. In this table, we also present the values of the
branching ratios of the semileptonic $\Xi_c \to B \ell \nu$ decays
obtained from other theoretical approaches, as well as the latest announced
experimental results. From a comparison of the predictions of the different
approaches, we see that our results are close to that of the ones given
in \cite{Ke:2021pxk} as well as the experimental measurements~\cite{Belle:2021crz,ALICE:2021bli}. On the other hand, the obtained branching ratios are slightly smaller than the results presented in \cite{Geng:2019bfz,Faustov:2019ddj,Zhang:2021oja} but larger than the values obtained in~\cite{Zhao:2021sje,Zhao:2018zcb}. However,  our results are considerably different for the results obtained in 
\cite{Azizi:2011mw} for the $\Xi_c \to \Xi \ell \nu$
decay, although they applied the same method as used in this work. This discrepancy can be explained as follows. The interpolating current of $\Xi_c$ baryon interacts not only with ground state positive parity baryons $J^P = (\frac{1}{2})^+$  but also with $J^P = \frac{1}{2}^-$ negative parity baryon which was neglected in~\cite{Azizi:2011mw}. Thus, the dispersion relation of $\Xi_c$ baryon is modified, and since the mass difference between these states is around 300 MeV, the results change considerably. 
\begin{turnpage}
  \begin{table*}
  \centering
  \renewcommand{\arraystretch}{1.4}
 \scriptsize
  \setlength{\tabcolsep}{4pt}
  \begin{tabular}{cccccccccccc}
    \toprule
  {\rm Decay~Channel} & {\rm Present~Work} & {\rm BELLE \cite{Belle:2021crz}}  & {\rm ALICE \cite{ALICE:2021bli}}  &  {\rm SU(3) \cite{Geng:2018plk}}   & {\rm SU(3) \cite{Geng:2019bfz}}  & {\rm RQM \cite{Faustov:2019ddj}} &  {\rm LATTICE \cite{Zhang:2021oja}} & {\rm 3PSR \cite{Zhao:2021sje}} & {\rm LCSR \cite{Azizi:2011mw}}    & {\rm LFQM \cite{Zhao:2018zcb}} & {\rm LF \cite{Ke:2021pxk}}   \\ 
    \midrule
  $\Xi_c^0 \to \Xi^- e^+ \nu_e $                 & $1.85 \pm 0.56$       &   $1.72 \pm 0.10 \pm 0.12 \pm 0.50 $ & $1.8 \pm 0.2$     & $ 4.87  \pm 1.74 $ & $ 2.4  \pm  0.3 $ & $ 2.38 $ & $ 2.38  \pm 0.30 \pm 0.33 $ & $ 1.45 \pm 0.31 $ & $ 7.26 \pm 2.54 $ & $ 1.354 $ & $ 1.72 \pm 0.35$   \\
  $\Xi_c^0 \to \Xi^- \mu^+ \nu_\mu $            & $   1.79 \pm 0.54        $ & $    1.71 \pm 0.17 \pm 0.13 \pm 0.50  $ & $ 1.8 \pm 0.2    $ & $ ---                            $ & $  2.4  \pm  0.3  $ & $ 2.31                           $ & $ 2.29  \pm 0.29 \pm 0.31       $ & $ 1.45 \pm 0.31                $ & $ 7.15 \pm 2.50 $ &  ---  &  \\
  $\Xi_c^+ \to \Xi^0 e^+ \nu_e                 $ & $   5.51 \pm 1.65      $ &    ---                                &  ---             & $ 3.38_{-2.26}^{+2.10}             $ & $ 9.8 \pm 1.1     $ & $  9.40                         $ & $ 7.18 \pm 0.90 \pm 0.98 $ &  ---  & $ 28.6 \pm 10 $ & $ 5.39 $ & $ 5.2 \pm 1.02$ \\
  $\Xi_c^+ \to \Xi^0 \mu^+ \nu_\mu             $ & $ 5.34 \pm 1.61         $ &  ---                                  &  ---            &  ---                             & $ 9.8 \pm 1.1                                    $ & $ 9.11 $ & $ 6.91 \pm 0.87 \pm 0.93   $ &  ---  & $ 28.2 \pm 9.9 $ &  ---  &  --- \\
 $ \Xi_c^+ \to \Lambda^0 e^+ \nu_e             $ & $ 0.092 \pm 0.028         $ &  ---                                   &  ---            &  ---                             & $ 0.166 \pm 0.018                                $ & $ 0.127 $ &  ---  &  ---  &  ---  & $ 0.082 $ &  --- \\                       %
 $ \Xi_c^+ \to \Lambda^0 \mu^+ \nu_\mu         $ & $ 0.089 \pm 0.027          $ &  ---                                   &  ---            &  ---                             &  ---                                             & $ 0.124 $ &  ---  &  ---  &  --- &  --- &  \\          
    \bottomrule
  \end{tabular}
\caption{The existing experimental and theoretical results on the branching ratios (in \%) of the
semileptonic $\Xi_c \rightarrow B l \nu$ decays.}
  \label{tab:tab4}
\end{table*}
\end{turnpage}

Our predictions on the branching ratios of $\Xi_c^+ \to \Lambda \ell \nu_\ell$ are also quite in agreement with the
results of \cite{Faustov:2019ddj} within the error. The predictions on the branching ratios can further be improved by more
precise determination of the input parameters appearing in DAs of the $\Xi$
and $\Lambda$ baryons, as well as taking into account ${\cal O}(\alpha_s)$
corrections.
\section{Conclusion}
The form factors of the semileptonic $\Xi_c \to \Xi (\Lambda) \ell \nu$ decays are studied in the framework of the light cone QCD sum rules method. In order to eliminate the contamination of the negative parity $\Xi_c^\ast$ baryon,
the combination of the sum rules obtained from different Lorentz structures is used.

Using the obtained results on the form factors and applying the helicity formalism, we also estimated the corresponding branching ratios of the considered decays. Moreover, our results on the branching ratios are compared with the predictions of the other approaches as well as with the experimental measurements.

The branching ratios of $\Xi_c \to \Xi \ell \nu$ decays has already been studied in various models like Relativistic Quark Model~\cite{Faustov:2019ddj}, LATTICE QCD~\cite{Zhang:2021oja}, 3-point sum rules~\cite{Zhao:2021sje}, Light Front Quark Models~\cite{Zhao:2018zcb,Ke:2021pxk}. Our calculations within the light cone sum rule showed that the resutls are in good agreement with the experimental measurements done by BELLE~\cite{Belle:2021crz} and ALICE~\cite{ALICE:2021bli} Collaborations.

The discrepancy between our finding and the results of \cite{Azizi:2011mw} in which the same method was used can be explained by taking into account the contributions of the $\Xi_c^\ast$ baryon that was neglected in \cite{Azizi:2011mw}.

Moreover, we also estimated the decay width of the CKM suppressed semileptonic $\Xi \rightarrow \Lambda l \nu$ decay within the light-cone sum rules. The obtained branching ratios are close the predictions of~\cite{Faustov:2019ddj} and the magnitude of the obtained value shows that it has potential to be measured in the future experiments

\bibliography{all.bib}

\begin{thebibliography}{21}%
\makeatletter
\providecommand \@ifxundefined [1]{%
 \@ifx{#1\undefined}
}%
\providecommand \@ifnum [1]{%
 \ifnum #1\expandafter \@firstoftwo
 \else \expandafter \@secondoftwo
 \fi
}%
\providecommand \@ifx [1]{%
 \ifx #1\expandafter \@firstoftwo
 \else \expandafter \@secondoftwo
 \fi
}%
\providecommand \natexlab [1]{#1}%
\providecommand \enquote  [1]{``#1''}%
\providecommand \bibnamefont  [1]{#1}%
\providecommand \bibfnamefont [1]{#1}%
\providecommand \citenamefont [1]{#1}%
\providecommand \href@noop [0]{\@secondoftwo}%
\providecommand \href [0]{\begingroup \@sanitize@url \@href}%
\providecommand \@href[1]{\@@startlink{#1}\@@href}%
\providecommand \@@href[1]{\endgroup#1\@@endlink}%
\providecommand \@sanitize@url [0]{\catcode `\\12\catcode `\$12\catcode
  `\&12\catcode `\#12\catcode `\^12\catcode `\_12\catcode `\%12\relax}%
\providecommand \@@startlink[1]{}%
\providecommand \@@endlink[0]{}%
\providecommand \url  [0]{\begingroup\@sanitize@url \@url }%
\providecommand \@url [1]{\endgroup\@href {#1}{\urlprefix }}%
\providecommand \urlprefix  [0]{URL }%
\providecommand \Eprint [0]{\href }%
\providecommand \doibase [0]{http://dx.doi.org/}%
\providecommand \selectlanguage [0]{\@gobble}%
\providecommand \bibinfo  [0]{\@secondoftwo}%
\providecommand \bibfield  [0]{\@secondoftwo}%
\providecommand \translation [1]{[#1]}%
\providecommand \BibitemOpen [0]{}%
\providecommand \bibitemStop [0]{}%
\providecommand \bibitemNoStop [0]{.\EOS\space}%
\providecommand \EOS [0]{\spacefactor3000\relax}%
\providecommand \BibitemShut  [1]{\csname bibitem#1\endcsname}%
\let\auto@bib@innerbib\@empty
\bibitem [{\citenamefont {Li}\ \emph {et~al.}(2021)\citenamefont {Li} \emph
  {et~al.}}]{Belle:2021crz}%
  \BibitemOpen
  \bibfield  {author} {\bibinfo {author} {\bibfnamefont {Y.~B.}\ \bibnamefont
  {Li}} \emph {et~al.} (\bibinfo {collaboration} {Belle}),\ }\bibfield  {title}
  {\enquote {\bibinfo {title} {{Measurements of the branching fractions of
  semileptonic decays $\Xi_{c}^{0} \to \Xi^{-} \ell^{+} \nu_{\ell}$ and
  asymmetry parameter of $\Xi_{c}^{0} \to \Xi^{-} \pi^{+}$ decay}},}\
  }\href@noop {} {\  (\bibinfo {year} {2021})},\ \Eprint
  {http://arxiv.org/abs/2103.06496} {arXiv:2103.06496 [hep-ex]} \BibitemShut
  {NoStop}%
\bibitem [{\citenamefont {Acharya}\ \emph {et~al.}(2021)\citenamefont {Acharya}
  \emph {et~al.}}]{ALICE:2021bli}%
  \BibitemOpen
  \bibfield  {author} {\bibinfo {author} {\bibfnamefont {Shreyasi}\
  \bibnamefont {Acharya}} \emph {et~al.} (\bibinfo {collaboration} {ALICE}),\
  }\bibfield  {title} {\enquote {\bibinfo {title} {{Measurement of the cross
  sections of $\Xi^0_{\rm c}$ and $\Xi^+_{\rm c}$ baryons and
  branching-fraction ratio BR($\Xi^0_{\rm c} \rightarrow \Xi^-{\rm e}^+\nu_{\rm
  e}$)/BR($\Xi^0_{\rm c} \rightarrow \Xi^-\pi^+$) in pp collisions at 13
  TeV}},}\ }\href@noop {} {\  (\bibinfo {year} {2021})},\ \Eprint
  {http://arxiv.org/abs/2105.05187} {arXiv:2105.05187 [nucl-ex]} \BibitemShut
  {NoStop}%
\bibitem [{\citenamefont {Geng}\ \emph {et~al.}(2018)\citenamefont {Geng},
  \citenamefont {Hsiao}, \citenamefont {Liu},\ and\ \citenamefont
  {Tsai}}]{Geng:2018plk}%
  \BibitemOpen
  \bibfield  {author} {\bibinfo {author} {\bibfnamefont {C.~Q.}\ \bibnamefont
  {Geng}}, \bibinfo {author} {\bibfnamefont {Y.~K.}\ \bibnamefont {Hsiao}},
  \bibinfo {author} {\bibfnamefont {Chia-Wei}\ \bibnamefont {Liu}}, \ and\
  \bibinfo {author} {\bibfnamefont {Tien-Hsueh}\ \bibnamefont {Tsai}},\
  }\bibfield  {title} {\enquote {\bibinfo {title} {{Antitriplet charmed baryon
  decays with SU(3) flavor symmetry}},}\ }\href {\doibase
  10.1103/PhysRevD.97.073006} {\bibfield  {journal} {\bibinfo  {journal} {Phys.
  Rev. D}\ }\textbf {\bibinfo {volume} {97}},\ \bibinfo {pages} {073006}
  (\bibinfo {year} {2018})},\ \Eprint {http://arxiv.org/abs/1801.03276}
  {arXiv:1801.03276 [hep-ph]} \BibitemShut {NoStop}%
\bibitem [{\citenamefont {Geng}\ \emph {et~al.}(2019)\citenamefont {Geng},
  \citenamefont {Liu}, \citenamefont {Tsai},\ and\ \citenamefont
  {Yeh}}]{Geng:2019bfz}%
  \BibitemOpen
  \bibfield  {author} {\bibinfo {author} {\bibfnamefont {Chao-Qiang}\
  \bibnamefont {Geng}}, \bibinfo {author} {\bibfnamefont {Chia-Wei}\
  \bibnamefont {Liu}}, \bibinfo {author} {\bibfnamefont {Tien-Hsueh}\
  \bibnamefont {Tsai}}, \ and\ \bibinfo {author} {\bibfnamefont {Shu-Wei}\
  \bibnamefont {Yeh}},\ }\bibfield  {title} {\enquote {\bibinfo {title}
  {{Semileptonic decays of anti-triplet charmed baryons}},}\ }\href {\doibase
  10.1016/j.physletb.2019.03.056} {\bibfield  {journal} {\bibinfo  {journal}
  {Phys. Lett. B}\ }\textbf {\bibinfo {volume} {792}},\ \bibinfo {pages}
  {214--218} (\bibinfo {year} {2019})},\ \Eprint
  {http://arxiv.org/abs/1901.05610} {arXiv:1901.05610 [hep-ph]} \BibitemShut
  {NoStop}%
\bibitem [{\citenamefont {Faustov}\ and\ \citenamefont
  {Galkin}(2019)}]{Faustov:2019ddj}%
  \BibitemOpen
  \bibfield  {author} {\bibinfo {author} {\bibfnamefont {R.~N.}\ \bibnamefont
  {Faustov}}\ and\ \bibinfo {author} {\bibfnamefont {V.~O.}\ \bibnamefont
  {Galkin}},\ }\bibfield  {title} {\enquote {\bibinfo {title} {{Semileptonic
  $\Xi_c$ baryon decays in the relativistic quark model}},}\ }\href {\doibase
  10.1140/epjc/s10052-019-7214-5} {\bibfield  {journal} {\bibinfo  {journal}
  {Eur. Phys. J. C}\ }\textbf {\bibinfo {volume} {79}},\ \bibinfo {pages} {695}
  (\bibinfo {year} {2019})},\ \Eprint {http://arxiv.org/abs/1905.08652}
  {arXiv:1905.08652 [hep-ph]} \BibitemShut {NoStop}%
\bibitem [{\citenamefont {Zhang}\ \emph {et~al.}(2021)\citenamefont {Zhang},
  \citenamefont {Hua}, \citenamefont {Huang}, \citenamefont {Li}, \citenamefont
  {Li}, \citenamefont {Lu}, \citenamefont {Sun}, \citenamefont {Sun},
  \citenamefont {Wang},\ and\ \citenamefont {Yang}}]{Zhang:2021oja}%
  \BibitemOpen
  \bibfield  {author} {\bibinfo {author} {\bibfnamefont {Qi-An}\ \bibnamefont
  {Zhang}}, \bibinfo {author} {\bibfnamefont {Jun}\ \bibnamefont {Hua}},
  \bibinfo {author} {\bibfnamefont {Fei}\ \bibnamefont {Huang}}, \bibinfo
  {author} {\bibfnamefont {Renbo}\ \bibnamefont {Li}}, \bibinfo {author}
  {\bibfnamefont {Yuanyuan}\ \bibnamefont {Li}}, \bibinfo {author}
  {\bibfnamefont {Cai-Dian}\ \bibnamefont {Lu}}, \bibinfo {author}
  {\bibfnamefont {Peng}\ \bibnamefont {Sun}}, \bibinfo {author} {\bibfnamefont
  {Wei}\ \bibnamefont {Sun}}, \bibinfo {author} {\bibfnamefont {Wei}\
  \bibnamefont {Wang}}, \ and\ \bibinfo {author} {\bibfnamefont {Yi-Bo}\
  \bibnamefont {Yang}},\ }\bibfield  {title} {\enquote {\bibinfo {title}
  {{$\Xi_c\to \Xi$ Form Factors and $\Xi_c\to \Xi \ell^+\nu_{\ell}$ Decay Rates
  From Lattice QCD}},}\ }\href@noop {} {\  (\bibinfo {year} {2021})},\ \Eprint
  {http://arxiv.org/abs/2103.07064} {arXiv:2103.07064 [hep-lat]} \BibitemShut
  {NoStop}%
\bibitem [{\citenamefont {Zhao}(2021)}]{Zhao:2021sje}%
  \BibitemOpen
  \bibfield  {author} {\bibinfo {author} {\bibfnamefont {Zhen-Xing}\
  \bibnamefont {Zhao}},\ }\bibfield  {title} {\enquote {\bibinfo {title}
  {{Semi-leptonic form factors of $\Xi_{c}\to\Xi$ in QCD sum rules}},}\
  }\href@noop {} {\  (\bibinfo {year} {2021})},\ \Eprint
  {http://arxiv.org/abs/2103.09436} {arXiv:2103.09436 [hep-ph]} \BibitemShut
  {NoStop}%
\bibitem [{\citenamefont {Azizi}\ \emph {et~al.}(2012)\citenamefont {Azizi},
  \citenamefont {Sarac},\ and\ \citenamefont {Sundu}}]{Azizi:2011mw}%
  \BibitemOpen
  \bibfield  {author} {\bibinfo {author} {\bibfnamefont {K.}~\bibnamefont
  {Azizi}}, \bibinfo {author} {\bibfnamefont {Y.}~\bibnamefont {Sarac}}, \ and\
  \bibinfo {author} {\bibfnamefont {H.}~\bibnamefont {Sundu}},\ }\bibfield
  {title} {\enquote {\bibinfo {title} {{Light cone QCD sum rules study of the
  semileptonic heavy $\Xi_{Q}$ and $\Xi'_{Q}$ transitions to $\Xi$ and $\Sigma
  $ baryons}},}\ }\href {\doibase 10.1140/epja/i2012-12002-1} {\bibfield
  {journal} {\bibinfo  {journal} {Eur. Phys. J. A}\ }\textbf {\bibinfo {volume}
  {48}},\ \bibinfo {pages} {2} (\bibinfo {year} {2012})},\ \Eprint
  {http://arxiv.org/abs/1107.5925} {arXiv:1107.5925 [hep-ph]} \BibitemShut
  {NoStop}%
\bibitem [{\citenamefont {Braun}\ \emph {et~al.}(2006)\citenamefont {Braun},
  \citenamefont {Lenz},\ and\ \citenamefont {Wittmann}}]{Braun:2006hz}%
  \BibitemOpen
  \bibfield  {author} {\bibinfo {author} {\bibfnamefont {V.~M.}\ \bibnamefont
  {Braun}}, \bibinfo {author} {\bibfnamefont {A.}~\bibnamefont {Lenz}}, \ and\
  \bibinfo {author} {\bibfnamefont {M.}~\bibnamefont {Wittmann}},\ }\bibfield
  {title} {\enquote {\bibinfo {title} {{Nucleon Form Factors in QCD}},}\ }\href
  {\doibase 10.1103/PhysRevD.73.094019} {\bibfield  {journal} {\bibinfo
  {journal} {Phys. Rev.}\ }\textbf {\bibinfo {volume} {D73}},\ \bibinfo {pages}
  {094019} (\bibinfo {year} {2006})},\ \Eprint
  {http://arxiv.org/abs/hep-ph/0604050} {arXiv:hep-ph/0604050 [hep-ph]}
  \BibitemShut {NoStop}%
\bibitem [{\citenamefont {Khodjamirian}\ \emph {et~al.}(2011)\citenamefont
  {Khodjamirian}, \citenamefont {Klein}, \citenamefont {Mannel},\ and\
  \citenamefont {Wang}}]{Khodjamirian:2011jp}%
  \BibitemOpen
  \bibfield  {author} {\bibinfo {author} {\bibfnamefont {A.}~\bibnamefont
  {Khodjamirian}}, \bibinfo {author} {\bibfnamefont {Ch.}\ \bibnamefont
  {Klein}}, \bibinfo {author} {\bibfnamefont {Th.}\ \bibnamefont {Mannel}}, \
  and\ \bibinfo {author} {\bibfnamefont {Y.~M.}\ \bibnamefont {Wang}},\
  }\bibfield  {title} {\enquote {\bibinfo {title} {{Form Factors and Strong
  Couplings of Heavy Baryons from QCD Light-Cone Sum Rules}},}\ }\href
  {\doibase 10.1007/JHEP09(2011)106} {\bibfield  {journal} {\bibinfo  {journal}
  {JHEP}\ }\textbf {\bibinfo {volume} {09}},\ \bibinfo {pages} {106} (\bibinfo
  {year} {2011})},\ \Eprint {http://arxiv.org/abs/1108.2971} {arXiv:1108.2971
  [hep-ph]} \BibitemShut {NoStop}%
\bibitem [{\citenamefont {Aliev}\ \emph {et~al.}(2015)\citenamefont {Aliev},
  \citenamefont {Azizi}, \citenamefont {Barakat},\ and\ \citenamefont
  {Savc\i{}}}]{Aliev:2015qea}%
  \BibitemOpen
  \bibfield  {author} {\bibinfo {author} {\bibfnamefont {T.~M.}\ \bibnamefont
  {Aliev}}, \bibinfo {author} {\bibfnamefont {K.}~\bibnamefont {Azizi}},
  \bibinfo {author} {\bibfnamefont {T.}~\bibnamefont {Barakat}}, \ and\
  \bibinfo {author} {\bibfnamefont {M.}~\bibnamefont {Savc\i{}}},\ }\bibfield
  {title} {\enquote {\bibinfo {title} {{Diagonal and transition magnetic
  moments of negative parity heavy baryons in QCD sum rules}},}\ }\href
  {\doibase 10.1103/PhysRevD.92.036004} {\bibfield  {journal} {\bibinfo
  {journal} {Phys. Rev. D}\ }\textbf {\bibinfo {volume} {92}},\ \bibinfo
  {pages} {036004} (\bibinfo {year} {2015})},\ \Eprint
  {http://arxiv.org/abs/1505.07977} {arXiv:1505.07977 [hep-ph]} \BibitemShut
  {NoStop}%
\bibitem [{\citenamefont {Bagan}\ \emph {et~al.}(1992)\citenamefont {Bagan},
  \citenamefont {Chabab}, \citenamefont {Dosch},\ and\ \citenamefont
  {Narison}}]{Bagan:1992tp}%
  \BibitemOpen
  \bibfield  {author} {\bibinfo {author} {\bibfnamefont {E.}~\bibnamefont
  {Bagan}}, \bibinfo {author} {\bibfnamefont {M.}~\bibnamefont {Chabab}},
  \bibinfo {author} {\bibfnamefont {Hans~Gunter}\ \bibnamefont {Dosch}}, \ and\
  \bibinfo {author} {\bibfnamefont {Stephan}\ \bibnamefont {Narison}},\
  }\bibfield  {title} {\enquote {\bibinfo {title} {{Spectra of heavy baryons
  from QCD spectral sum rules}},}\ }\href {\doibase
  10.1016/0370-2693(92)91896-H} {\bibfield  {journal} {\bibinfo  {journal}
  {Phys. Lett. B}\ }\textbf {\bibinfo {volume} {287}},\ \bibinfo {pages}
  {176--178} (\bibinfo {year} {1992})}\BibitemShut {NoStop}%
\bibitem [{\citenamefont {Liu}\ and\ \citenamefont
  {Huang}(2009{\natexlab{a}})}]{Liu:2009uc}%
  \BibitemOpen
  \bibfield  {author} {\bibinfo {author} {\bibfnamefont {Yong-Lu}\ \bibnamefont
  {Liu}}\ and\ \bibinfo {author} {\bibfnamefont {Ming-Qiu}\ \bibnamefont
  {Huang}},\ }\bibfield  {title} {\enquote {\bibinfo {title} {{Light-cone
  Distribution Amplitudes of $\Xi$ and their Applications}},}\ }\href {\doibase
  10.1103/PhysRevD.80.055015} {\bibfield  {journal} {\bibinfo  {journal} {Phys.
  Rev.}\ }\textbf {\bibinfo {volume} {D80}},\ \bibinfo {pages} {055015}
  (\bibinfo {year} {2009}{\natexlab{a}})},\ \Eprint
  {http://arxiv.org/abs/0909.0372} {arXiv:0909.0372 [hep-ph]} \BibitemShut
  {NoStop}%
\bibitem [{\citenamefont {Liu}\ and\ \citenamefont
  {Huang}(2009{\natexlab{b}})}]{Liu:2008yg}%
  \BibitemOpen
  \bibfield  {author} {\bibinfo {author} {\bibfnamefont {Yong-Lu}\ \bibnamefont
  {Liu}}\ and\ \bibinfo {author} {\bibfnamefont {Ming-Qiu}\ \bibnamefont
  {Huang}},\ }\bibfield  {title} {\enquote {\bibinfo {title} {{Distribution
  amplitudes of $\Sigma$ and $\Lambda $ and their electromagnetic form
  factors}},}\ }\href {\doibase 10.1016/j.nuclphysa.2009.02.003} {\bibfield
  {journal} {\bibinfo  {journal} {Nucl. Phys.}\ }\textbf {\bibinfo {volume}
  {A821}},\ \bibinfo {pages} {80--105} (\bibinfo {year}
  {2009}{\natexlab{b}})},\ \Eprint {http://arxiv.org/abs/0811.1812}
  {arXiv:0811.1812 [hep-ph]} \BibitemShut {NoStop}%
\bibitem [{\citenamefont {Wein}\ and\ \citenamefont
  {Sch{\"a}fer}(2015)}]{Wein:2015oqa}%
  \BibitemOpen
  \bibfield  {author} {\bibinfo {author} {\bibfnamefont {Philipp}\ \bibnamefont
  {Wein}}\ and\ \bibinfo {author} {\bibfnamefont {Andreas}\ \bibnamefont
  {Sch{\"a}fer}},\ }\bibfield  {title} {\enquote {\bibinfo {title}
  {{Model-independent calculation of SU(3)$_{f}$ violation in baryon octet
  light-cone distribution amplitudes}},}\ }\href {\doibase
  10.1007/JHEP05(2015)073} {\bibfield  {journal} {\bibinfo  {journal} {JHEP}\
  }\textbf {\bibinfo {volume} {05}},\ \bibinfo {pages} {073} (\bibinfo {year}
  {2015})},\ \Eprint {http://arxiv.org/abs/1501.07218} {arXiv:1501.07218
  [hep-ph]} \BibitemShut {NoStop}%
\bibitem [{\citenamefont {Zyla}\ \emph {et~al.}(2020)\citenamefont {Zyla} \emph
  {et~al.}}]{ParticleDataGroup:2020ssz}%
  \BibitemOpen
  \bibfield  {author} {\bibinfo {author} {\bibfnamefont {P.~A.}\ \bibnamefont
  {Zyla}} \emph {et~al.} (\bibinfo {collaboration} {Particle Data Group}),\
  }\bibfield  {title} {\enquote {\bibinfo {title} {{Review of Particle
  Physics}},}\ }\href {\doibase 10.1093/ptep/ptaa104} {\bibfield  {journal}
  {\bibinfo  {journal} {PTEP}\ }\textbf {\bibinfo {volume} {2020}},\ \bibinfo
  {pages} {083C01} (\bibinfo {year} {2020})}\BibitemShut {NoStop}%
\bibitem [{\citenamefont {Chetyrkin}\ \emph {et~al.}(2009)\citenamefont
  {Chetyrkin}, \citenamefont {Kuhn}, \citenamefont {Maier}, \citenamefont
  {Maierhofer}, \citenamefont {Marquard}, \citenamefont {Steinhauser},\ and\
  \citenamefont {Sturm}}]{Chetyrkin:2009fv}%
  \BibitemOpen
  \bibfield  {author} {\bibinfo {author} {\bibfnamefont {K.~G.}\ \bibnamefont
  {Chetyrkin}}, \bibinfo {author} {\bibfnamefont {J.~H.}\ \bibnamefont {Kuhn}},
  \bibinfo {author} {\bibfnamefont {A.}~\bibnamefont {Maier}}, \bibinfo
  {author} {\bibfnamefont {P.}~\bibnamefont {Maierhofer}}, \bibinfo {author}
  {\bibfnamefont {P.}~\bibnamefont {Marquard}}, \bibinfo {author}
  {\bibfnamefont {M.}~\bibnamefont {Steinhauser}}, \ and\ \bibinfo {author}
  {\bibfnamefont {C.}~\bibnamefont {Sturm}},\ }\bibfield  {title} {\enquote
  {\bibinfo {title} {{Charm and Bottom Quark Masses: An Update}},}\ }\href
  {\doibase 10.1103/PhysRevD.80.074010} {\bibfield  {journal} {\bibinfo
  {journal} {Phys. Rev. D}\ }\textbf {\bibinfo {volume} {80}},\ \bibinfo
  {pages} {074010} (\bibinfo {year} {2009})},\ \Eprint
  {http://arxiv.org/abs/0907.2110} {arXiv:0907.2110 [hep-ph]} \BibitemShut
  {NoStop}%
\bibitem [{\citenamefont {Bourrely}\ \emph {et~al.}(2009)\citenamefont
  {Bourrely}, \citenamefont {Caprini},\ and\ \citenamefont
  {Lellouch}}]{Bourrely:2008za}%
  \BibitemOpen
  \bibfield  {author} {\bibinfo {author} {\bibfnamefont {Claude}\ \bibnamefont
  {Bourrely}}, \bibinfo {author} {\bibfnamefont {Irinel}\ \bibnamefont
  {Caprini}}, \ and\ \bibinfo {author} {\bibfnamefont {Laurent}\ \bibnamefont
  {Lellouch}},\ }\bibfield  {title} {\enquote {\bibinfo {title}
  {{Model-independent description of $B \rightarrow \pi l \nu$ decays and a
  determination of $|V_{ub}|$}},}\ }\href {\doibase 10.1103/PhysRevD.82.099902}
  {\bibfield  {journal} {\bibinfo  {journal} {Phys. Rev. D}\ }\textbf {\bibinfo
  {volume} {79}},\ \bibinfo {pages} {013008} (\bibinfo {year} {2009})},\
  \bibinfo {note} {[Erratum: Phys.Rev.D 82, 099902 (2010)]},\ \Eprint
  {http://arxiv.org/abs/0807.2722} {arXiv:0807.2722 [hep-ph]} \BibitemShut
  {NoStop}%
\bibitem [{\citenamefont {Gutsche}\ \emph {et~al.}(2015)\citenamefont
  {Gutsche}, \citenamefont {Ivanov}, \citenamefont {K\"orner}, \citenamefont
  {Lyubovitskij}, \citenamefont {Santorelli},\ and\ \citenamefont
  {Habyl}}]{Gutsche:2015mxa}%
  \BibitemOpen
  \bibfield  {author} {\bibinfo {author} {\bibfnamefont {Thomas}\ \bibnamefont
  {Gutsche}}, \bibinfo {author} {\bibfnamefont {Mikhail~A.}\ \bibnamefont
  {Ivanov}}, \bibinfo {author} {\bibfnamefont {J\"urgen~G.}\ \bibnamefont
  {K\"orner}}, \bibinfo {author} {\bibfnamefont {Valery~E.}\ \bibnamefont
  {Lyubovitskij}}, \bibinfo {author} {\bibfnamefont {Pietro}\ \bibnamefont
  {Santorelli}}, \ and\ \bibinfo {author} {\bibfnamefont {Nurgul}\ \bibnamefont
  {Habyl}},\ }\bibfield  {title} {\enquote {\bibinfo {title} {{Semileptonic
  decay $\Lambda_b \to \Lambda_c + \tau^- + \bar{\nu_\tau}$ in the covariant
  confined quark model}},}\ }\href {\doibase 10.1103/PhysRevD.91.074001}
  {\bibfield  {journal} {\bibinfo  {journal} {Phys. Rev. D}\ }\textbf {\bibinfo
  {volume} {91}},\ \bibinfo {pages} {074001} (\bibinfo {year} {2015})},\
  \bibinfo {note} {[Erratum: Phys.Rev.D 91, 119907 (2015)]},\ \Eprint
  {http://arxiv.org/abs/1502.04864} {arXiv:1502.04864 [hep-ph]} \BibitemShut
  {NoStop}%
\bibitem [{\citenamefont {Ke}\ \emph {et~al.}(2021)\citenamefont {Ke},
  \citenamefont {Kang}, \citenamefont {Liu},\ and\ \citenamefont
  {Li}}]{Ke:2021pxk}%
  \BibitemOpen
  \bibfield  {author} {\bibinfo {author} {\bibfnamefont {Hong-Wei}\
  \bibnamefont {Ke}}, \bibinfo {author} {\bibfnamefont {Qing-Qing}\
  \bibnamefont {Kang}}, \bibinfo {author} {\bibfnamefont {Xiao-Hai}\
  \bibnamefont {Liu}}, \ and\ \bibinfo {author} {\bibfnamefont {Xue-Qian}\
  \bibnamefont {Li}},\ }\bibfield  {title} {\enquote {\bibinfo {title} {{The
  weak decays of $\Xi^{(')}_{c}\to\Xi$ in the light-front quark model}},}\
  }\href@noop {} {\  (\bibinfo {year} {2021})},\ \Eprint
  {http://arxiv.org/abs/2106.07013} {arXiv:2106.07013 [hep-ph]} \BibitemShut
  {NoStop}%
\bibitem [{\citenamefont {Zhao}(2018)}]{Zhao:2018zcb}%
  \BibitemOpen
  \bibfield  {author} {\bibinfo {author} {\bibfnamefont {Zhen-Xing}\
  \bibnamefont {Zhao}},\ }\bibfield  {title} {\enquote {\bibinfo {title} {{Weak
  decays of heavy baryons in the light-front approach}},}\ }\href {\doibase
  10.1088/1674-1137/42/9/093101} {\bibfield  {journal} {\bibinfo  {journal}
  {Chin. Phys. C}\ }\textbf {\bibinfo {volume} {42}},\ \bibinfo {pages}
  {093101} (\bibinfo {year} {2018})},\ \Eprint
  {http://arxiv.org/abs/1803.02292} {arXiv:1803.02292 [hep-ph]} \BibitemShut
  {NoStop}%
\end{thebibliography}%

\end{document}